\PassOptionsToPackage{unicode}{hyperref}
\PassOptionsToPackage{hyphens}{url}
\PassOptionsToPackage{dvipsnames,svgnames,x11names}{xcolor}
\documentclass[
]{article}

\usepackage{amsmath,amssymb}
\usepackage{iftex}
\ifPDFTeX
  \usepackage[T1]{fontenc}
  \usepackage[utf8]{inputenc}
  \usepackage{textcomp} 
\else 
  \usepackage{unicode-math}
  \defaultfontfeatures{Scale=MatchLowercase}
  \defaultfontfeatures[\rmfamily]{Ligatures=TeX,Scale=1}
\fi
\usepackage{lmodern}
\ifPDFTeX\else  
\fi
\IfFileExists{upquote.sty}{\usepackage{upquote}}{}
\IfFileExists{microtype.sty}{
  \usepackage[]{microtype}
  \UseMicrotypeSet[protrusion]{basicmath} 
}{}
\makeatletter
\@ifundefined{KOMAClassName}{
  \IfFileExists{parskip.sty}{%
    \usepackage{parskip}
  }{
    \setlength{\parindent}{0pt}
    \setlength{\parskip}{6pt plus 2pt minus 1pt}}
}{
  \KOMAoptions{parskip=half}}
\makeatother
\usepackage{xcolor}
\ifx\paragraph\undefined\else
  \let\oldparagraph\paragraph
  \renewcommand{\paragraph}[1]{\oldparagraph{#1}\mbox{}}
\fi
\ifx\subparagraph\undefined\else
  \let\oldsubparagraph\subparagraph
  \renewcommand{\subparagraph}[1]{\oldsubparagraph{#1}\mbox{}}
\fi

\usepackage{color}
\usepackage{fancyvrb}

\DefineVerbatimEnvironment{Highlighting}{Verbatim}{commandchars=\\\{\}}
\usepackage{framed}
\definecolor{shadecolor}{RGB}{241,243,245}
\newenvironment{Shaded}{\begin{snugshade}}{\end{snugshade}}

\newcommand{\AttributeTok}[1]{\textcolor[rgb]{0.40,0.45,0.13}{#1}}

\newcommand{\CommentTok}[1]{\textcolor[rgb]{0.37,0.37,0.37}{#1}}

\newcommand{\ConstantTok}[1]{\textcolor[rgb]{0.56,0.35,0.01}{#1}}
\newcommand{\ControlFlowTok}[1]{\textcolor[rgb]{0.00,0.23,0.31}{#1}}

\newcommand{\DecValTok}[1]{\textcolor[rgb]{0.68,0.00,0.00}{#1}}

\newcommand{\FloatTok}[1]{\textcolor[rgb]{0.68,0.00,0.00}{#1}}
\newcommand{\FunctionTok}[1]{\textcolor[rgb]{0.28,0.35,0.67}{#1}}

\newcommand{\NormalTok}[1]{\textcolor[rgb]{0.00,0.23,0.31}{#1}}

\newcommand{\OtherTok}[1]{\textcolor[rgb]{0.00,0.23,0.31}{#1}}

\newcommand{\SpecialCharTok}[1]{\textcolor[rgb]{0.37,0.37,0.37}{#1}}

\newcommand{\StringTok}[1]{\textcolor[rgb]{0.13,0.47,0.30}{#1}}

\providecommand{\tightlist}{%
  \setlength{\itemsep}{0pt}\setlength{\parskip}{0pt}}\usepackage{longtable,booktabs,array}
\usepackage{calc} 
\usepackage{etoolbox}
\makeatletter
\patchcmd\longtable{\par}{\if@noskipsec\mbox{}\fi\par}{}{}
\makeatother
\IfFileExists{footnotehyper.sty}{\usepackage{footnotehyper}}{\usepackage{footnote}}
\makesavenoteenv{longtable}
\usepackage{graphicx}
\makeatletter
\def\maxwidth{\ifdim\Gin@nat@width>\linewidth\linewidth\else\Gin@nat@width\fi}
\def\maxheight{\ifdim\Gin@nat@height>\textheight\textheight\else\Gin@nat@height\fi}
\makeatother
\setkeys{Gin}{width=\maxwidth,height=\maxheight,keepaspectratio}
\makeatletter
\def\fps@figure{htbp}
\makeatother
\NewDocumentCommand\citeproctext{}{}

\makeatletter
 \let\@cite@ofmt\@firstofone
 \def\@biblabel#1{}
 \def\@cite#1#2{{#1\if@tempswa , #2\fi}}
\makeatother
\newlength{\cslhangindent}
\newlength{\csllabelwidth}
\newenvironment{CSLReferences}[2] 
 {\begin{list}{}{%
  \setlength{\itemindent}{0pt}
  \setlength{\leftmargin}{0pt}
  \setlength{\parsep}{0pt}
  \ifodd #1
   \setlength{\leftmargin}{\cslhangindent}
   \setlength{\itemindent}{-1\cslhangindent}
  \fi
  \setlength{\itemsep}{#2\baselineskip}}}
 {\end{list}}
\usepackage{calc}

\usepackage{arxiv}
\usepackage{orcidlink}
\usepackage{amsmath}
\usepackage[T1]{fontenc}
\usepackage{colortbl}
\usepackage{bera}
\usepackage{xcolor}
\usepackage{hyperref}
\usepackage[utf8]{inputenc}
\def\tightlist{}

\makeatletter
\@ifpackageloaded{tcolorbox}{}{\usepackage[skins,breakable]{tcolorbox}}
\@ifpackageloaded{fontawesome5}{}{\usepackage{fontawesome5}}
\definecolor{quarto-callout-color}{HTML}{909090}
\definecolor{quarto-callout-note-color}{HTML}{0758E5}
\definecolor{quarto-callout-important-color}{HTML}{CC1914}
\definecolor{quarto-callout-warning-color}{HTML}{EB9113}
\definecolor{quarto-callout-tip-color}{HTML}{00A047}
\definecolor{quarto-callout-caution-color}{HTML}{FC5300}
\definecolor{quarto-callout-color-frame}{HTML}{acacac}
\definecolor{quarto-callout-note-color-frame}{HTML}{4582ec}
\definecolor{quarto-callout-important-color-frame}{HTML}{d9534f}
\definecolor{quarto-callout-warning-color-frame}{HTML}{f0ad4e}
\definecolor{quarto-callout-tip-color-frame}{HTML}{02b875}
\definecolor{quarto-callout-caution-color-frame}{HTML}{fd7e14}
\makeatother
\makeatletter
\@ifpackageloaded{caption}{}{\usepackage{caption}}
\AtBeginDocument{%
\ifdefined\contentsname
  \renewcommand*\contentsname{Table of contents}
\else
  \newcommand\contentsname{Table of contents}
\fi
\ifdefined\listfigurename
  \renewcommand*\listfigurename{List of Figures}
\else
  \newcommand\listfigurename{List of Figures}
\fi
\ifdefined\listtablename
  \renewcommand*\listtablename{List of Tables}
\else
  \newcommand\listtablename{List of Tables}
\fi
\ifdefined\figurename
  \renewcommand*\figurename{Figure}
\else
  \newcommand\figurename{Figure}
\fi
\ifdefined\tablename
  \renewcommand*\tablename{Table}
\else
  \newcommand\tablename{Table}
\fi
}
\@ifpackageloaded{float}{}{\usepackage{float}}
\floatstyle{ruled}
\@ifundefined{c@chapter}{\newfloat{codelisting}{h}{lop}}{\newfloat{codelisting}{h}{lop}[chapter]}
\floatname{codelisting}{Listing}

\makeatother
\makeatletter
\@ifpackageloaded{caption}{}{\usepackage{caption}}
\@ifpackageloaded{subcaption}{}{\usepackage{subcaption}}
\makeatother
\makeatletter
\@ifpackageloaded{tikz}{}{\usepackage{tikz}}
\makeatother
        \newcommand*\circled[1]{\tikz[baseline=(char.base)]{
          \node[shape=circle,draw,inner sep=1pt] (char) {{\scriptsize#1}};}}  
                  
\ifLuaTeX
  \usepackage{selnolig}  
\fi
\IfFileExists{bookmark.sty}{\usepackage{bookmark}}{\usepackage{hyperref}}
\IfFileExists{xurl.sty}{\usepackage{xurl}}{} 
\hypersetup{
  pdftitle={edibble: An R package to encapsulate elements of experimental designs for better planning, management and workflow},
  pdfauthor={Emi Tanaka},
  pdfkeywords={grammar of experimental designs, design of
experiments, comparative experiments, interface design, grammarware},
  colorlinks=true,
  linkcolor={blue},
  filecolor={Maroon},
  citecolor={Blue},
  urlcolor={Blue},
  pdfcreator={LaTeX via pandoc}}

\newcommand{\runninghead}{A Preprint }
\title{{edibble}: An {R} package to encapsulate elements of experimental
designs for better planning, management and workflow}
\author{
\textbf{Emi Tanaka}~\orcidlink{0000-0002-1455-259X}\\Biological Data
Science Institute\\Australian National
University\\Canberra\\\href{mailto:emi.tanaka@anu.edu.au}{emi.tanaka@anu.edu.au}}
\date{}
\begin{document}
\maketitle
\begin{abstract}
I present an {R} package called {edibble} that facilitates the design of
experiments by encapsulating elements of the experiment in a series of
composable functions. This package is an interpretation of ``the grammar
of experimental designs'' by Tanaka (2023) in the {R} programming
language. The main features of the {edibble} package are demonstrated,
illustrating how it can be used to create a wide array of experimental
designs. The implemented system aims to encourage cognitive thinking for
holistic planning and data management of experiments in a streamlined
workflow. This workflow can increase the inherent value of experimental
data by reducing potential errors or noise with careful preplanning, as
well as, ensuring fit-for-purpose analysis of experimental data.
\end{abstract}
{\bfseries \emph Keywords}
\def\sep{\textbullet\ }
grammar of experimental designs \sep design of
experiments \sep comparative experiments \sep interface design \sep 
grammarware

\section{Introduction}\label{introduction}

The critical role of data collection is well captured in the expression
``garbage in, garbage out'' -- in other words, if the collected data is
rubbish, then no analysis, however complex it may be, can make something
out of it. Experiments offer the highest degree of control in the method
of data collection and are of a critical importance for validating or
investigating hypotheses across numerous fields (e.g., agriculture,
biology, chemistry, business, marketing, engineering). A proper design
of experiments is critical to ensure that the desired inference is valid
and efficient based on the experimental data. Conducting experiments is
usually expensive; thus improper designs are at best inefficient use of
resources and, at worst, a complete waste of resources. Given such high
stakes, it is clear that improving the practice of experimental design
can translate to large gains by increasing the inherent value of the
experimental data.

However, more holistically, experiments involve multiple steps that
extend beyond generating an experimental design table or layout. An
appropriate experimental design cannot be generated without context and
subject matter expertise to identify the appropriate experimental
factors (Hahn 1984; Steinberg and Hunter 1984). Furthermore, the
experimental design is useless if it is not carried out as intended or
if there are issues with data entry. Generating an experimental design
should not be seen as an isolated activity, but rather viewed as an
interconnected activity within the life of the whole experiment. Even a
fit-for-purpose analysis of experimental data is not an independent
activity void of the knowledge of the experimental design. Naturally, we
can develop systems for constructing experimental designs that support
an integrated approach for the entire lifecycle of an experiment.

Many experimental design software programs do not offer an integrated
approach. Rather, the software tightly integrates the algorithmic aspect
into a given experimental structure, thus losing the flexibility to
specify other structures (Tanaka and Amaliah 2022). Unfortunately, this
tight integration causes friction when considering alternative
algorithms and requires users to rebuild the specification from scratch
for another software solution. In addition, because the processes to
specify the design are generally not modular, users must process all
aspects of the experimental design simultaneously, burdening their
cognitive load. As an alternative framework to specify experimental
designs, Tanaka (2023) presented a computational framework, called ``the
grammar of experimental designs'', which allows for intermediate
constructs of experimental designs. This framework adopts a
process-based approach to support the gathering of required information
for the entire experiment in a modular manner.

In Tanaka (2023), an experimental design is treated as a mutable object
that progressively builds to the final design by a series of functions
that modify a targetted element of the experimental design. This system
is applicable to a broad range of experimental designs with fixed levels
for each experimental factor. The system can be understood as an analogy
to grammar in linguistics; a vast array of experimental designs
(sentences) is expressed by combining key functions (words) through the
shared understanding of the rules (grammar). In this system, an
intermediate construct of an experimental design is represented as a
pair of directed acyclic graphs (DAGs), where the nodes of one DAG
represent the experimental factors, whereas the nodes of the other DAG
represent the levels of those factors. Users explicitly define the
experimental factors, their roles (such as treatment, unit or record),
and relationships between the nodes. These purposeful manipulations of
the DAGs are invoked by a small collection of object methods. The names
of these methods (e.g., ``set units'' and ``assign treatments'') are
semantically aligned to raise conscious awareness of the user (and the
reader) to the elements of the experimental design. The {edibble}
package in the {R} language (R Core Team 2023) was used to demonstrate
the utility of the framework in Tanaka (2023), however, little
explanation was given to the usage of the package.

This paper extensively demonstrates the main features of the {edibble}
package. Section~\ref{sec-usage} provides an overview of its usage,
Section~\ref{sec-define} describes various methods for defining the
experimental structure. Section~\ref{sec-examples} presents additional
examples based on real experiments. Section~\ref{sec-contrast} contrasts
the existing systems. Finally, we conclude the paper with a discussion
in Section~\ref{sec-conclude}.

\section{Usage overview}\label{sec-usage}

The {edibble} package is available on the Comprehensive R Archive
Network (\url{https://cran.r-project.org/}) with the latest developments
available on GitHub at \url{https://github.com/emitanaka/edibble}. The
code presented in this paper is based on version 1.1.0 of the {edibble}
package.

\begin{figure}

\centering{

\includegraphics{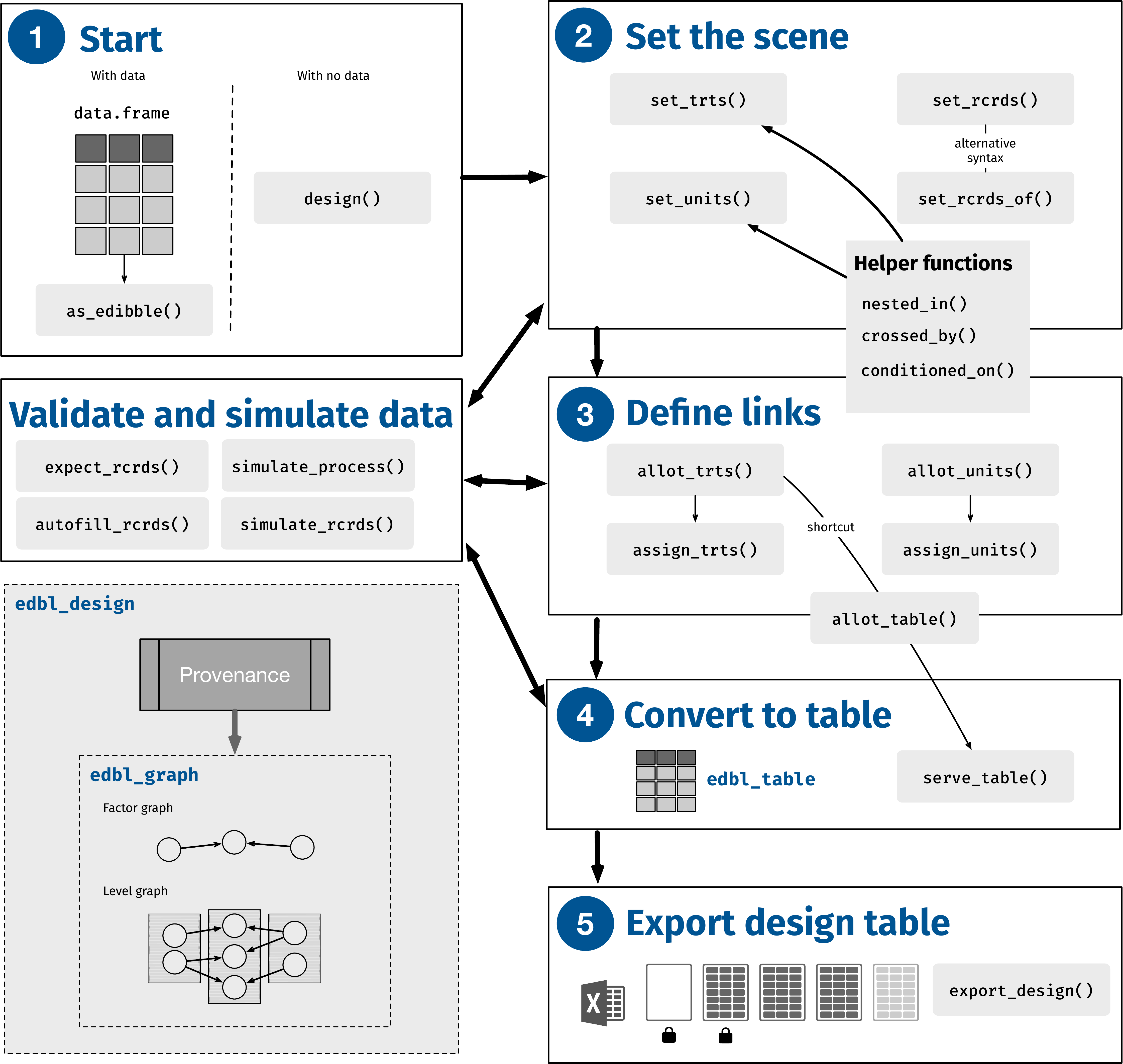}

}

\caption{\label{fig-workflow}The general workflow for using {edibble}.
The design starts with either no data or with data. We then set the
scene by defining the experimental factors with its roles and assign
links between the factors and levels. The information is stored as an
{edbl\_graph} object enclosed in an {edbl\_design} object that also
encloses the {Provenance} object that peforms low-level manipulations of
the graphs. Once the full experimental structure is defined, then the
table can be produced and exported. The record factors can be validated
or simulated. The usage details of the functions are presented in
Section~\ref{sec-define}.}

\end{figure}

The ultimate aim of the {edibble} package is to produce an experimental
design table. The name reflects this aim, with {edibble} standing for
\textbf{e}xperimental \textbf{d}esign t\textbf{ibble}, where {tibble} is
a special type of {data.frame} from {tibble} package (Müller and Wickham
2023b). The key functions to achieve this aim are illustrated in
Figure~\ref{fig-workflow}. An example shown next demonstrates a quick
overview of its usage.

Consider Example 1, derived from the example in Bailey (2008) for a
quick demonstration of the {edibble} package. The main aim of this
experiment is to determine the best feed type for calves to gain weight.

\begin{tcolorbox}[enhanced jigsaw, breakable, opacitybacktitle=0.6, leftrule=.75mm, colback=white, opacityback=0, colbacktitle=quarto-callout-note-color!10!white, title=\textcolor{quarto-callout-note-color}{\faInfo}\hspace{0.5em}{Example 1: Calf feeding}, bottomrule=.15mm, arc=.35mm, coltitle=black, bottomtitle=1mm, toprule=.15mm, toptitle=1mm, left=2mm, titlerule=0mm, colframe=quarto-callout-note-color-frame, rightrule=.15mm]

There were 8 pens, with 10 calves in each pen. The experimenter was
interested in comparing the effects of the four types of feed on the
calves' weights. Each calf was individually weighed. The four feeds were
combinations of two types of hay, which were put into the pen, and two
types of anti-scour treatments, which were administered individually to
each calf.

\end{tcolorbox}

To specify this experiment in {edibble}, we composed it using a series
of functions as shown below. The pipe (\texttt{\%\textgreater{}\%})
function, imported from {magrittr} (Bache and Wickham 2022), allows for
a series of nested operations and can be substituted with the native
pipe (\texttt{\textbar{}\textgreater{}}) available from {R} version
4.1.0 onwards. This style of coding would be familiar to users of
{tidyverse} (Wickham et al. 2019).

\phantomsection\label{annotated-cell-1}%
\begin{Shaded}
\begin{Highlighting}[]
\FunctionTok{library}\NormalTok{(edibble)}
\NormalTok{calf }\OtherTok{\textless{}{-}} \FunctionTok{design}\NormalTok{(}\StringTok{"Calf feeding"}\NormalTok{) }\SpecialCharTok{\%\textgreater{}\%}\hspace*{\fill}\NormalTok{\circled{1}}
  \FunctionTok{set\_units}\NormalTok{(}\AttributeTok{pen =} \DecValTok{8}\NormalTok{,}\hspace*{\fill}\NormalTok{\circled{2}}
            \AttributeTok{calf =} \FunctionTok{nested\_in}\NormalTok{(pen, }\DecValTok{10}\NormalTok{)) }\SpecialCharTok{\%\textgreater{}\%}
  \FunctionTok{set\_rcrds}\NormalTok{(}\AttributeTok{weight =}\NormalTok{ calf) }\SpecialCharTok{\%\textgreater{}\%}\hspace*{\fill}\NormalTok{\circled{3}}
  \CommentTok{\# set\_trts(feed = 4) \%\textgreater{}\% }\hspace*{\fill}\NormalTok{\circled{4}}
  \FunctionTok{set\_trts}\NormalTok{(}\AttributeTok{hay =} \DecValTok{2}\NormalTok{,}\hspace*{\fill}\NormalTok{\circled{5}}
           \AttributeTok{antiscour =} \DecValTok{2}\NormalTok{) }\SpecialCharTok{\%\textgreater{}\%}
  \FunctionTok{allot\_table}\NormalTok{(hay }\SpecialCharTok{\textasciitilde{}}\NormalTok{ pen,}\hspace*{\fill}\NormalTok{\circled{6}}
\NormalTok{              antiscour }\SpecialCharTok{\textasciitilde{}}\NormalTok{ calf)}
\end{Highlighting}
\end{Shaded}

\begin{description}
\tightlist
\item[\circled{1}]
We begin the design specification by initiating the {edbl\_design}
object using {design}. An optional title is provided, which is encoded
and persistently displayed at various outputs (e.g., print or export of
the object).
\item[\circled{2}]
As described in the example, we had 8 pens with 10 calves in each pen.
This part of the code specifies the unit factors ``pen'' and ``calf''.
The right hand side shows the number of levels as a single numerical
value. The function {nested\_in} is a helper function that encodes the
calf is nested in pen.
\item[\circled{3}]
As each calf is individually weighed, we set the records such that the
weight is recorded for each calf (\texttt{weight\ =\ calf}). Here, the
right hand side has to be a unit factor that has been previously
defined.
\item[\circled{4}]
We are told that we are comparing four types of feed; hence, we set the
treatments to \texttt{feed\ =\ 4}, but we realise later that this is not
the entire description. We comment this line out and instead write our
understanding in the next lines. You can, of course, choose to delete
this line, but it can be useful to keep a record of your process.\\
\item[\circled{5}]
Here, we write our new understanding that the feed is composed of two
types of hay and two types of anti-scour treatments.
\item[\circled{6}]
In the final step, {allot\_table} is a short hand for {allot\_trts},
{assign\_trts}, and {serve\_table}; as such, multiple processes occur in
this call. The relationship between the units and treatments is
specified: hay type is alloted and assigned randomly to the pen, and
anti-scour treatment type is alloted and assigned randomly to the calf
within the pen. The design object is then translated into a table.
\end{description}

Printing the object above shows the table below, in which each
experimental factor is translated into a column.

\begin{Shaded}
\begin{Highlighting}[]
\NormalTok{calf}
\end{Highlighting}
\end{Shaded}

\begin{verbatim}
# Calf feeding 
# An edibble: 80 x 5
     pen    calf  weight    hay  antiscour
  <U(8)> <U(80)> <R(80)> <T(2)>     <T(2)>
   <chr>   <chr>   <dbl>  <chr>      <chr>
1   pen1  calf01       o   hay1 antiscour1
2   pen1  calf02       o   hay1 antiscour2
3   pen1  calf03       o   hay1 antiscour2
4   pen1  calf04       o   hay1 antiscour1
5   pen1  calf05       o   hay1 antiscour1
6   pen1  calf06       o   hay1 antiscour2
# i 74 more rows
\end{verbatim}

\begin{figure}

\centering{

\includegraphics{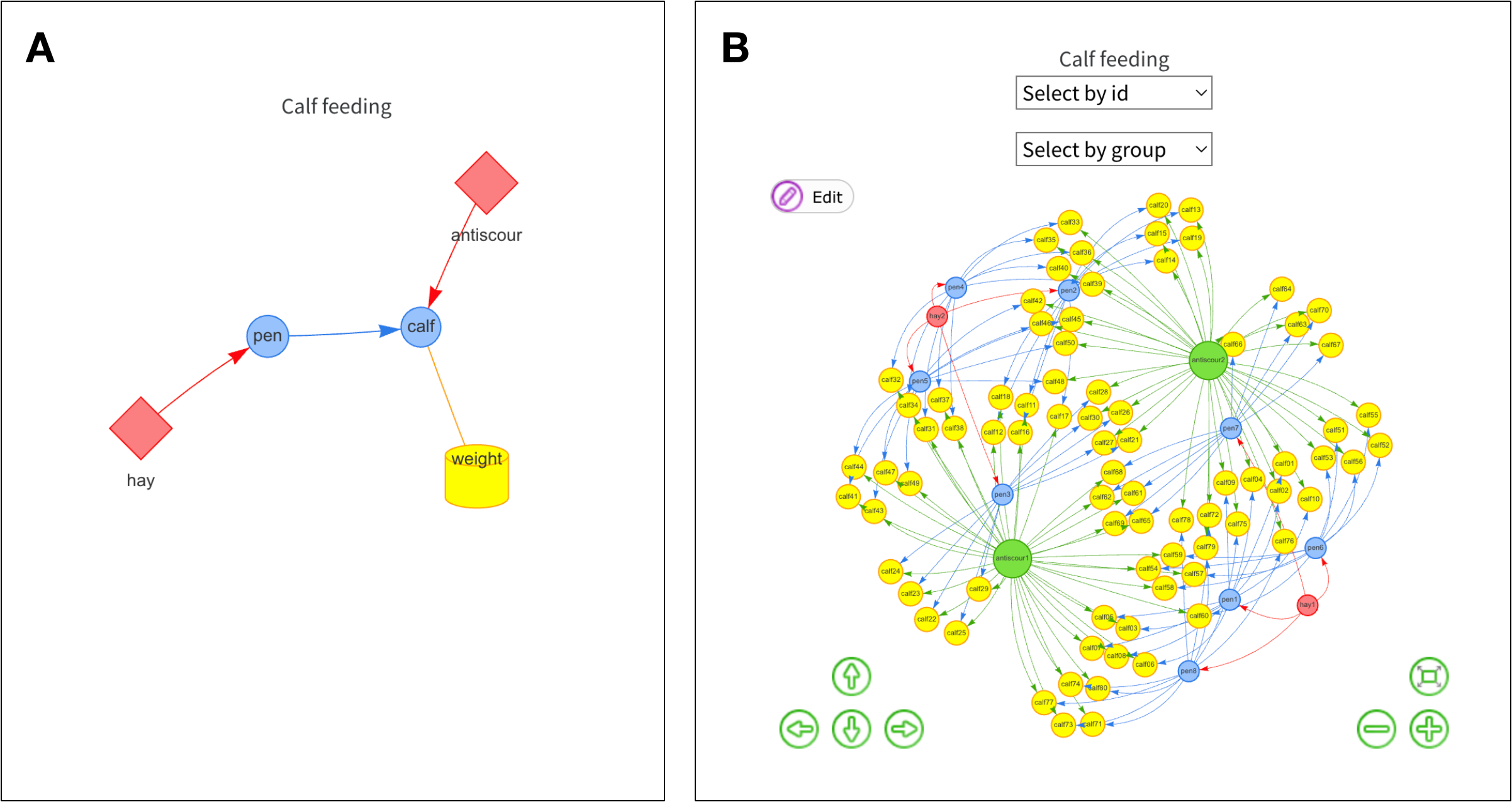}

}

\caption{\label{fig-graphs}(A) Factor graph for the calf experiment. The
shape and color of the nodes correspond to the role of the factor (red
diamond = treatment, blue circle = unit, yellow three-dimensional
cylinder = record). The labels correspond to the name of the factor and
arrows show the relationship of the factors. Here we have the hay
treatment factor alloted to pen and antiscour treatment factor alloted
to calf. The weight is recorded for each calf. (B) A level graph for the
calf experiment. The colors represent different experimental factors.
The dropdown menu can be used to select a particular node or group of
nodes that belong to the same experimental factor. Selecting a node
highlights all connecting nodes, which can be useful for verifying the
links between nodes.}

\end{figure}

The resultant object also contains the {edbl\_design} object (hereafter
referred plainly as the design object). The user may also plot the
internal graphs using the {plot} function. By default, this shows the
factor graph as an interactive plot by internally using the {visNetwork}
package (Almende B.V. and Contributors and Thieurmel 2022). The level
graph can be seen by adding the argument as \texttt{which\ =\ "levels"}
in the {plot} function. The static plots of the factor and level graphs
are shown in Figure~\ref{fig-graphs}. The level graph can contain many
nodes since every level of each experimental factor, except the record
factors, is shown.

\begin{Shaded}
\begin{Highlighting}[]
\FunctionTok{plot}\NormalTok{(calf) }\CommentTok{\# factor graph is default}
\FunctionTok{plot}\NormalTok{(calf, }\AttributeTok{which =} \StringTok{"levels"}\NormalTok{) }\CommentTok{\# level graph}
\end{Highlighting}
\end{Shaded}

In experiments, there would be a response of interest. A response is
often not required to generate an experimental layout. It is generally
assumed that the smallest unit in the experimental design table is the
observational unit (although not always so). In {edibble}, users can
optionally specify their intention of what to record (e.g., responses)
on which particular unit factor, as was done above, where the intention
to capture the \texttt{weight} of each \texttt{calf} is specified in the
function {set\_rcrds}. We can further add the expectations of the values
for the \texttt{weight} factor. For example, below, we use
{expect\_rcrds} to encode a data validation that the weight should be a
numeric value between 0 and 10.

\begin{Shaded}
\begin{Highlighting}[]
\NormalTok{calfr }\OtherTok{\textless{}{-}}\NormalTok{ calf }\SpecialCharTok{\%\textgreater{}\%} 
  \FunctionTok{expect\_rcrds}\NormalTok{(weight }\SpecialCharTok{\textgreater{}} \DecValTok{0}\NormalTok{, weight }\SpecialCharTok{\textless{}} \DecValTok{10}\NormalTok{)}
\end{Highlighting}
\end{Shaded}

The benefit of encoding the data validation rule is that it is now
interoperable. Two functionalities in {edibble} take advantage of this
encoding. The first is when exporting the design through
{export\_design}. This outputs an Excel file with the column ``weight''
in the data sheet for calves. The cells in this column are empty, with
the intention that the data collector enters the data in this Excel
sheet. These cells have the data validation embedded such that if the
entry is not a numeric value between 0 and 10, it will result in an
error in the data entry. See more details in Section~\ref{sec-expect}.

The other functionality in {edibble} that takes advantage of the data
validation encoding is when simulating data for record factor(s). There
are two approaches to simulating data in {edibble}: one that requires
users to extensively describe the simulation process
({simulate\_process} and then {simulate\_rcrds}, and the other that
automatically writes the simulation scheme for you while ensuring to
generate valid values: {autofill\_rcrds}. When users describe the
simulation data on their own, it is possible to violate the valid
values. In this case, the values are censored (by default to missing
values) in {simulate\_rcrds}. More details on the simulation
capabilities are presented in Section~\ref{sec-simulate}.

\section{Defining structure}\label{sec-define}

An \emph{experimental structure} always consists of a \emph{unit
structure}; in other words, an experiment cannot be defined without
units. If the experiment is comparative, then it must consist of a
\emph{treatment structure} in addition to the mapping between the unit
and treatment factors. This mapping can be defined on two levels: a
high-level mapping between factors (usually for humans to understand the
broad picture), and a low-level mapping between factor levels (usually
assigned algorithmically).

There are three roles for the experimental factors that can be encoded
in {edibble}: treatment, unit, and record (see more details in Tanaka
2023). The functions, {set\_trts}, {set\_units}, and {set\_rcrds} are
used to encode the treatment, unit, and record factors, respectively,
into the design object. The calls to these functions are generally
associative, that is, the order does not matter nor does the number of
calls to it matter (e.g., you do not need to define all units in one
call and instead call on {set\_units} repeatedly). The exception is when
the new factor is dependent on a previously defined factor; in this
case, the dependent factor must be defined later.

Once the factors are defined, the relationship between them and their
levels must be defined. The high-level grammar restricts the users in
the type of relationships that can be defined (e.g., you cannot assign
treatments to record factors), although low-level tweaks to the grammar
(meant for developers) can bypass these restrictions (not presented in
this paper).

The main functions in {edibble} are explained in detail In the following
subsections. In this section, we assume that the experimental aim is to
find the best wheat variety from a wheat field trial.

\subsection{Initialisation}\label{initialisation}

\begin{Shaded}
\begin{Highlighting}[]
\FunctionTok{library}\NormalTok{(edibble)}
\end{Highlighting}
\end{Shaded}

A new design constructed using {edibble} must start by initialising the
design object. An optional title of the design may be provided as input.
This information persists as metadata in the object and is displayed in
various places (e.g., print output and exported files).

\subsubsection{With data}\label{with-data}

If you have existing data, you can use it as a base to build the design
object. Below, we convert the data \texttt{gilmour.serpentine} in
{agridat} (Wright 2022) into an {edbl\_table} object.

\begin{Shaded}
\begin{Highlighting}[]
\FunctionTok{as\_edibble}\NormalTok{(agridat}\SpecialCharTok{::}\NormalTok{gilmour.serpentine, }
           \AttributeTok{.title =} \StringTok{"Wheat Experiment"}\NormalTok{, }
           \AttributeTok{.units =} \FunctionTok{c}\NormalTok{(col, row, rep), }\AttributeTok{.trts =}\NormalTok{ gen) }
\end{Highlighting}
\end{Shaded}

\begin{verbatim}
# Wheat Experiment 
# An edibble: 330 x 5
      col     row    rep      gen yield
  <U(15)> <U(22)> <U(3)> <T(107)>      
    <int>   <int>  <fct>    <fct> <int>
1       1       1      1       4    483
2       1       2      1       10   526
3       1       3      1       15   557
4       1       4      1       17   564
5       1       5      1       21   498
6       1       6      1       32   510
# i 324 more rows
\end{verbatim}

The main benefit of converting an existing experimental data into an
{edbl\_table} format is that other functionalities offered by {edibble}
can be used.

For the remainder of the paper, we assume experiments have no prior data
that is used directly in the experimental design (as is commonly the
case when you are conducting an experiment from scratch).

\subsubsection{With no data}\label{with-no-data}

When you have no data, you start by simply initialising the design
object.

\begin{Shaded}
\begin{Highlighting}[]
\FunctionTok{design}\NormalTok{(}\StringTok{"Wheat field trial"}\NormalTok{)}
\end{Highlighting}
\end{Shaded}

At this point, there is nothing particularly interesting. The design
object requires the user to define the experimental factor(s) as
described next.

\subsection{Units}\label{units}

At minimum, the design requires units to be defined via {set\_units}. In
the code below, we initialise a new design object and then set a unit
called ``site'' with 4 levels. The left hand side (LHS) and the right
hand side (RHS) of the function input correspond to the factor name and
the corresponding value, respectively. Here, the value is a single
integer that denotes the number of levels of the factor. Note that the
LHS can be any arbitrary (preferably syntactically valid) name.
Selecting a name that succinctly describes the factor is recommended.
Acronyms should be avoided where reasonable. We assign this design
object to the variable called \texttt{demo}.

\begin{Shaded}
\begin{Highlighting}[]
\NormalTok{demo }\OtherTok{\textless{}{-}} \FunctionTok{design}\NormalTok{(}\StringTok{"Demo for defining units"}\NormalTok{) }\SpecialCharTok{\%\textgreater{}\%} 
  \FunctionTok{set\_units}\NormalTok{(}\AttributeTok{site =} \DecValTok{4}\NormalTok{)}
\end{Highlighting}
\end{Shaded}

At this point, the design is in a \emph{graph form}. The print of this
object shows a prettified tree that displays the title of the
experiment, the factors, and their corresponding number of levels.
Notice the root in this tree output corresponds to the title given in
the object initialisation.

\begin{Shaded}
\begin{Highlighting}[]
\NormalTok{demo}
\end{Highlighting}
\end{Shaded}

\begin{verbatim}
Demo for defining units
\-site (4 levels)
\end{verbatim}

To obtain the \emph{design table}, you must call on {serve\_table} to
signal that you wish the object to be transformed into the \emph{tabular
form}. The transformation for \texttt{demo} is shown below, where the
output is a type of {tibble} with one column (the ``site'' factor), four
rows (corresponding to the four levels in the site), and the entries
corresponding to the actual levels of the factor (name derived as
``site1'', ``site2'', ``site3'', and ``site4'' here). The first line of
the print output is decorated with the title of the design object, which
acts as a persistent reminder of the initial input. The row just under
the header shows the role of the factor denoted by the upper case letter
(here, U = unit) with the number of levels in that factor displayed via
{pillar} (Müller and Wickham 2023a) with methods encoded for a custom
vector class using the {vctrs} package (Wickham, Henry, and Vaughan
2023). If the number of levels exceed a thousand, then the number is
shown with an SI prefix rounded to the closest digit corresponding to
the SI prefix form (e.g., 1000 is shown as 1k and 1800 is shown as
\textasciitilde2k). The row that follows shows the class of the factor
(e.g., character or numeric).

\begin{Shaded}
\begin{Highlighting}[]
\FunctionTok{serve\_table}\NormalTok{(demo)}
\end{Highlighting}
\end{Shaded}

\begin{verbatim}
# Demo for defining units 
# An edibble: 4 x 1
    site
  <U(4)>
   <chr>
1  site1
2  site2
3  site3
4  site4
\end{verbatim}

If particular names are desired for the levels, then the RHS value can
be replaced with a vector like below where the levels are named
``Narrabri'', ``Horsham'', ``Parkes'' and ``Roseworthy''.

\begin{Shaded}
\begin{Highlighting}[]
\FunctionTok{design}\NormalTok{(}\StringTok{"Character vector input demo"}\NormalTok{) }\SpecialCharTok{\%\textgreater{}\%} 
  \FunctionTok{set\_units}\NormalTok{(}\AttributeTok{site =} \FunctionTok{c}\NormalTok{(}\StringTok{"Narrabri"}\NormalTok{, }\StringTok{"Horsham"}\NormalTok{, }\StringTok{"Parkes"}\NormalTok{, }\StringTok{"Roseworthy"}\NormalTok{)) }\SpecialCharTok{\%\textgreater{}\%} 
  \FunctionTok{serve\_table}\NormalTok{()}
\end{Highlighting}
\end{Shaded}

\begin{verbatim}
# Character vector input demo 
# An edibble: 4 x 1
        site
      <U(4)>
       <chr>
1 Narrabri  
2 Horsham   
3 Parkes    
4 Roseworthy
\end{verbatim}

The RHS value in theory be any vector. Below the input is a numeric
vector, and the corresponding output will be a {data.frame} with a
numeric column.

\begin{Shaded}
\begin{Highlighting}[]
\FunctionTok{design}\NormalTok{(}\StringTok{"Numeric vector input demo"}\NormalTok{) }\SpecialCharTok{\%\textgreater{}\%} 
  \FunctionTok{set\_units}\NormalTok{(}\AttributeTok{site =} \FunctionTok{c}\NormalTok{(}\DecValTok{1}\NormalTok{, }\DecValTok{2}\NormalTok{, }\DecValTok{3}\NormalTok{, }\DecValTok{4}\NormalTok{)) }\SpecialCharTok{\%\textgreater{}\%} 
  \FunctionTok{serve\_table}\NormalTok{()}
\end{Highlighting}
\end{Shaded}

\begin{verbatim}
# Numeric vector input demo 
# An edibble: 4 x 1
    site
  <U(4)>
   <dbl>
1      1
2      2
3      3
4      4
\end{verbatim}

In the instance that you do want to enter a single level with a numeric
value, this can be specified using {lvls} on the RHS.

\begin{Shaded}
\begin{Highlighting}[]
\FunctionTok{design}\NormalTok{(}\StringTok{"Single numeric level demo"}\NormalTok{) }\SpecialCharTok{\%\textgreater{}\%} 
  \FunctionTok{set\_units}\NormalTok{(}\AttributeTok{site =} \FunctionTok{lvls}\NormalTok{(}\DecValTok{4}\NormalTok{)) }\SpecialCharTok{\%\textgreater{}\%} 
  \FunctionTok{serve\_table}\NormalTok{()}
\end{Highlighting}
\end{Shaded}

\begin{verbatim}
# Single numeric level demo 
# An edibble: 1 x 1
    site
  <U(1)>
   <dbl>
1      4
\end{verbatim}

\subsubsection{Multiple units}\label{multiple-units}

We can add more unit factors to this study. Suppose that we have 72
plots. We append another call to {set\_units} to encode this
information.

\begin{Shaded}
\begin{Highlighting}[]
\NormalTok{demo2 }\OtherTok{\textless{}{-}}\NormalTok{ demo }\SpecialCharTok{\%\textgreater{}\%} 
  \FunctionTok{set\_units}\NormalTok{(}\AttributeTok{plot =} \DecValTok{72}\NormalTok{)}
\end{Highlighting}
\end{Shaded}

However, we did not defined the relationship between \texttt{site} and
\texttt{plot}; so it fails to convert to the tabular form.

\begin{Shaded}
\begin{Highlighting}[]
\FunctionTok{serve\_table}\NormalTok{(demo2)}
\end{Highlighting}
\end{Shaded}

\begin{verbatim}
Error in `serve_table()`:
! The graph cannot be converted to a table format.
\end{verbatim}

The relationship between unit factors can be defined concurrently when
defining the unit factors using helper functions. One of these helper
functions is demonstrated next.

\subsubsection{Nested units}\label{sec-nested}

Given that we have a wheat trial, we imagine that the site corresponds
to the locations, and each location would have its own plots. The
experimenter tells you that each site contains 18 plots. This nesting
structure can be defined by using the helper function {nested\_in}. With
this relationship specified, the graph can be reconciled into a tabular
format, as shown below.

\begin{Shaded}
\begin{Highlighting}[]
\NormalTok{demo }\SpecialCharTok{\%\textgreater{}\%} 
  \FunctionTok{set\_units}\NormalTok{(}\AttributeTok{plot =} \FunctionTok{nested\_in}\NormalTok{(site, }\DecValTok{18}\NormalTok{)) }\SpecialCharTok{\%\textgreater{}\%} 
  \FunctionTok{serve\_table}\NormalTok{()}
\end{Highlighting}
\end{Shaded}

\begin{verbatim}
# Demo for defining units 
# An edibble: 72 x 2
    site    plot
  <U(4)> <U(72)>
   <chr>   <chr>
1  site1  plot01
2  site1  plot02
3  site1  plot03
4  site1  plot04
5  site1  plot05
6  site1  plot06
# i 66 more rows
\end{verbatim}

In the above situation, the relationship between unit factors have to be
apriori known, but there are situations in which the relationship may
become cognizant only after defining the unit factors. In these
situations, users can define the relationships using the functions
{allot\_units} and {assign\_units} to add the edges between the relevant
unit nodes in the factor and level graphs, respectively (see
Section~\ref{sec-allot} and Section~\ref{sec-assign} for more details of
these functions).

\begin{Shaded}
\begin{Highlighting}[]
\NormalTok{demo2 }\SpecialCharTok{\%\textgreater{}\%} 
  \FunctionTok{allot\_units}\NormalTok{(site }\SpecialCharTok{\textasciitilde{}}\NormalTok{ plot) }\SpecialCharTok{\%\textgreater{}\%} 
  \FunctionTok{assign\_units}\NormalTok{(}\AttributeTok{order =} \StringTok{"systematic{-}fastest"}\NormalTok{) }\SpecialCharTok{\%\textgreater{}\%} 
  \FunctionTok{serve\_table}\NormalTok{()}
\end{Highlighting}
\end{Shaded}

\begin{verbatim}
# Demo for defining units 
# An edibble: 72 x 2
    site    plot
  <U(4)> <U(72)>
   <chr>   <chr>
1  site1  plot01
2  site2  plot02
3  site3  plot03
4  site4  plot04
5  site1  plot05
6  site2  plot06
# i 66 more rows
\end{verbatim}

The code above specifies the nested relationship of \texttt{plot} to
\texttt{site}, with the assignment of levels performed systematically.
The systematic allocation of \texttt{site} levels to \texttt{plot} is
done so that the \texttt{site} levels vary the fastest, which is not the
same systematic ordering as before. If the same result as before is
desirable, users can define \texttt{order\ =\ "systematic-slowest"},
which offers a systematic assignment where the same levels are close
together.

\subsubsection{Crossed units}\label{sec-crossed}

Crop field trials are often laid out in rectangular arrays. The
experimenter confirms this by alerting to us that each site has plots
laid out in a rectangular array with 6 rows and 3 columns. We can define
crossing structures using {crossed\_by}.

\begin{Shaded}
\begin{Highlighting}[]
\FunctionTok{design}\NormalTok{(}\StringTok{"Crossed experiment"}\NormalTok{) }\SpecialCharTok{\%\textgreater{}\%} 
  \FunctionTok{set\_units}\NormalTok{(}\AttributeTok{row =} \DecValTok{6}\NormalTok{,}
            \AttributeTok{col =} \DecValTok{3}\NormalTok{,}
            \AttributeTok{plot =} \FunctionTok{crossed\_by}\NormalTok{(row, col)) }\SpecialCharTok{\%\textgreater{}\%} 
  \FunctionTok{serve\_table}\NormalTok{()}
\end{Highlighting}
\end{Shaded}

\begin{verbatim}
# Crossed experiment 
# An edibble: 18 x 3
     row    col    plot
  <U(6)> <U(3)> <U(18)>
   <chr>  <chr>   <chr>
1   row1   col1  plot01
2   row2   col1  plot02
3   row3   col1  plot03
4   row4   col1  plot04
5   row5   col1  plot05
6   row6   col1  plot06
# i 12 more rows
\end{verbatim}

The above table does not contain information on the site. For this, we
need to combine the nesting and crossing structures, as shown next.

\subsubsection{Complex unit structures}\label{complex-unit-structures}

Now, suppose that there are four sites (Narrabri, Horsham, Parkes, and
Roseworthy), and the 18 plots at each site are laid out in a rectangular
array of 3 rows and 6 columns. We begin by specifying the site (the
highest hierarchy in this structure). The dimensions of the rows and
columns are specified for each site (3 rows and 6 columns). The plot is
a result of crossing the row and column within each site.

\begin{Shaded}
\begin{Highlighting}[]
\NormalTok{complex }\OtherTok{\textless{}{-}} \FunctionTok{design}\NormalTok{(}\StringTok{"Complex structure"}\NormalTok{) }\SpecialCharTok{\%\textgreater{}\%} 
  \FunctionTok{set\_units}\NormalTok{(}\AttributeTok{site =} \FunctionTok{c}\NormalTok{(}\StringTok{"Narrabri"}\NormalTok{, }\StringTok{"Horsham"}\NormalTok{, }\StringTok{"Parkes"}\NormalTok{, }\StringTok{"Roseworthy"}\NormalTok{),}
            \AttributeTok{col =} \FunctionTok{nested\_in}\NormalTok{(site, }\DecValTok{6}\NormalTok{),}
            \AttributeTok{row =} \FunctionTok{nested\_in}\NormalTok{(site, }\DecValTok{3}\NormalTok{),}
            \AttributeTok{plot =} \FunctionTok{nested\_in}\NormalTok{(site, }\FunctionTok{crossed\_by}\NormalTok{(row, col))) }

\FunctionTok{serve\_table}\NormalTok{(complex)}
\end{Highlighting}
\end{Shaded}

\begin{verbatim}
# Complex structure 
# An edibble: 72 x 4
       site     col     row    plot
     <U(4)> <U(24)> <U(12)> <U(72)>
      <chr>   <chr>   <chr>   <chr>
 1 Narrabri   col01   row01  plot01
 2 Narrabri   col01   row02  plot02
 3 Narrabri   col01   row03  plot03
 4 Narrabri   col02   row01  plot04
 5 Narrabri   col02   row02  plot05
 6 Narrabri   col02   row03  plot06
 7 Narrabri   col03   row01  plot07
 8 Narrabri   col03   row02  plot08
 9 Narrabri   col03   row03  plot09
10 Narrabri   col04   row01  plot10
11 Narrabri   col04   row02  plot11
12 Narrabri   col04   row03  plot12
13 Narrabri   col05   row01  plot13
14 Narrabri   col05   row02  plot14
15 Narrabri   col05   row03  plot15
16 Narrabri   col06   row01  plot16
17 Narrabri   col06   row02  plot17
18 Narrabri   col06   row03  plot18
19 Horsham    col07   row04  plot19
20 Horsham    col07   row05  plot20
# i 52 more rows
\end{verbatim}

You may realise that the labels for the rows do not start with ``row1''
for Horsham. The default output displays distinct labels for the unit
levels that are actually distinct. This safeguards for instances where
the relationship between factors is lost, and the analyst will have to
guess what units may be nested or crossed. However, nested labels may
still be desirable. You can select the factors to show the nested labels
by naming these factors as arguments for the \texttt{label\_nested} in
{serve\_table} (below shows the nesting labels for \texttt{row} and
\texttt{col} -- notice \texttt{plot} still shows the distinct labels).

\begin{Shaded}
\begin{Highlighting}[]
\FunctionTok{serve\_table}\NormalTok{(complex, }\AttributeTok{label\_nested =} \FunctionTok{c}\NormalTok{(row, col))}
\end{Highlighting}
\end{Shaded}

\begin{verbatim}
# Complex structure 
# An edibble: 72 x 4
       site     col     row    plot
     <U(4)> <U(24)> <U(12)> <U(72)>
      <chr>   <chr>   <chr>   <chr>
 1 Narrabri    col1    row1  plot01
 2 Narrabri    col1    row2  plot02
 3 Narrabri    col1    row3  plot03
 4 Narrabri    col2    row1  plot04
 5 Narrabri    col2    row2  plot05
 6 Narrabri    col2    row3  plot06
 7 Narrabri    col3    row1  plot07
 8 Narrabri    col3    row2  plot08
 9 Narrabri    col3    row3  plot09
10 Narrabri    col4    row1  plot10
11 Narrabri    col4    row2  plot11
12 Narrabri    col4    row3  plot12
13 Narrabri    col5    row1  plot13
14 Narrabri    col5    row2  plot14
15 Narrabri    col5    row3  plot15
16 Narrabri    col6    row1  plot16
17 Narrabri    col6    row2  plot17
18 Narrabri    col6    row3  plot18
19 Horsham     col1    row1  plot19
20 Horsham     col1    row2  plot20
# i 52 more rows
\end{verbatim}

You later find that the dimensions of Narrabri and Roseworthy are
larger. The experimenter tells you that there are in fact 9 columns
available, and therefore 27 plots at Narrabri and Roseworthy. The number
of columns can be modified according to each site, as below, where
\texttt{col} is defined to have 9 levels at Narrabri and Roseworthy but
6 levels elsewhere.

\begin{Shaded}
\begin{Highlighting}[]
\NormalTok{complexd }\OtherTok{\textless{}{-}} \FunctionTok{design}\NormalTok{(}\StringTok{"Complex structure with different dimensions"}\NormalTok{) }\SpecialCharTok{\%\textgreater{}\%} 
  \FunctionTok{set\_units}\NormalTok{(}\AttributeTok{site =} \FunctionTok{c}\NormalTok{(}\StringTok{"Narrabri"}\NormalTok{, }\StringTok{"Horsham"}\NormalTok{, }\StringTok{"Parkes"}\NormalTok{, }\StringTok{"Roseworthy"}\NormalTok{),}
             \AttributeTok{col =} \FunctionTok{nested\_in}\NormalTok{(site, }
                      \FunctionTok{c}\NormalTok{(}\StringTok{"Narrabri"}\NormalTok{, }\StringTok{"Roseworthy"}\NormalTok{) }\SpecialCharTok{\textasciitilde{}} \DecValTok{9}\NormalTok{,}
\NormalTok{                                                . }\SpecialCharTok{\textasciitilde{}} \DecValTok{6}\NormalTok{),}
             \AttributeTok{row =} \FunctionTok{nested\_in}\NormalTok{(site, }\DecValTok{3}\NormalTok{),}
            \AttributeTok{plot =} \FunctionTok{nested\_in}\NormalTok{(site, }\FunctionTok{crossed\_by}\NormalTok{(row, col))) }

\NormalTok{complextab }\OtherTok{\textless{}{-}} \FunctionTok{serve\_table}\NormalTok{(complexd, }\AttributeTok{label\_nested =} \FunctionTok{everything}\NormalTok{())}
\FunctionTok{table}\NormalTok{(complextab}\SpecialCharTok{$}\NormalTok{site)}
\end{Highlighting}
\end{Shaded}

\begin{verbatim}

   Horsham   Narrabri     Parkes Roseworthy 
        18         27         18         27 
\end{verbatim}

You can see above that there are indeed nine additional plots at
Narrabri and Roseworthy. The argument for \texttt{label\_nested}
supports {tidyselect} (Henry and Wickham 2022) approach for selecting
factors.

\subsection{Treatments}\label{sec-trt}

Defining treatment factors is only necessary when designing a
comparative experiment. The treatment factors can be set similar to the
unit factors using {set\_trts}. Below, we define an experiment with
three treatment factors: variety (a or b), fertilizer (A or B), and
amount of fertilizer (0.5, 1, or 2 t/ha).

\begin{Shaded}
\begin{Highlighting}[]
\NormalTok{factrt }\OtherTok{\textless{}{-}} \FunctionTok{design}\NormalTok{(}\StringTok{"Factorial treatment"}\NormalTok{) }\SpecialCharTok{\%\textgreater{}\%} 
  \FunctionTok{set\_trts}\NormalTok{(}\AttributeTok{variety =} \FunctionTok{c}\NormalTok{(}\StringTok{"a"}\NormalTok{, }\StringTok{"b"}\NormalTok{),}
           \AttributeTok{fertilizer =} \FunctionTok{c}\NormalTok{(}\StringTok{"A"}\NormalTok{, }\StringTok{"B"}\NormalTok{),}
           \AttributeTok{amount =} \FunctionTok{c}\NormalTok{(}\FloatTok{0.5}\NormalTok{, }\DecValTok{1}\NormalTok{, }\DecValTok{2}\NormalTok{)) }
\end{Highlighting}
\end{Shaded}

The links between treatment factors need not be explicitly defined. It
is automatically assumed that treatment factors are crossed (i.e., the
resulting treatment is the combination of all treatment factors) with
the full set of treatments shown via {trts\_table}. For the above
experiment, there are a total of 12 treatments with the levels given
below.

\begin{Shaded}
\begin{Highlighting}[]
\FunctionTok{trts\_table}\NormalTok{(factrt)}
\end{Highlighting}
\end{Shaded}

\begin{verbatim}
# A tibble: 12 x 3
   variety fertilizer amount
   <chr>   <chr>       <dbl>
 1 a       A             0.5
 2 b       A             0.5
 3 a       B             0.5
 4 b       B             0.5
 5 a       A             1  
 6 b       A             1  
 7 a       B             1  
 8 b       B             1  
 9 a       A             2  
10 b       A             2  
11 a       B             2  
12 b       B             2  
\end{verbatim}

The \texttt{factrt} cannot be served as an {edbl\_table} object, since
there are no units defined in this experiment and how these treatments
are administered to the units.

\subsubsection{Conditional treatments}\label{sec-conditioned}

In some experiments, certain treatment factors are dependent on another
treatment factor. A common example is when the dose or amount of a
treatment factor is also a treatment factor. In the field trial example,
we can have a case in which we administer no fertilizer to a plot. In
this case, there is no point crossing with different \texttt{amount}s;
in fact, the amount of no fertilizer should always be 0. We can specify
this conditional treatment structure by describing this relationship
using the helper function, {conditioned\_on}, as below. The ``.'' in the
LHS is a shorthand to mean all levels, except for those specified
previously.

\begin{Shaded}
\begin{Highlighting}[]
\NormalTok{factrtc }\OtherTok{\textless{}{-}} \FunctionTok{design}\NormalTok{(}\StringTok{"Factorial treatment with control"}\NormalTok{) }\SpecialCharTok{\%\textgreater{}\%} 
  \FunctionTok{set\_trts}\NormalTok{(}\AttributeTok{variety =} \FunctionTok{c}\NormalTok{(}\StringTok{"a"}\NormalTok{, }\StringTok{"b"}\NormalTok{),}
           \AttributeTok{fertilizer =} \FunctionTok{c}\NormalTok{(}\StringTok{"none"}\NormalTok{, }\StringTok{"A"}\NormalTok{, }\StringTok{"B"}\NormalTok{),}
           \AttributeTok{amount =} \FunctionTok{conditioned\_on}\NormalTok{(fertilizer,}
                                    \StringTok{"none"} \SpecialCharTok{\textasciitilde{}} \DecValTok{0}\NormalTok{,}
\NormalTok{                                         . }\SpecialCharTok{\textasciitilde{}} \FunctionTok{c}\NormalTok{(}\FloatTok{0.5}\NormalTok{, }\DecValTok{1}\NormalTok{, }\DecValTok{2}\NormalTok{)))}
\end{Highlighting}
\end{Shaded}

We can see below that the variety is crossed with other factors, as
expected, but the amount is conditional on the fertilizer.

\begin{Shaded}
\begin{Highlighting}[]
\FunctionTok{trts\_table}\NormalTok{(factrtc)}
\end{Highlighting}
\end{Shaded}

\begin{verbatim}
# A tibble: 14 x 3
   variety fertilizer amount
   <chr>   <chr>       <dbl>
 1 a       none          0  
 2 b       none          0  
 3 a       A             0.5
 4 b       A             0.5
 5 a       A             1  
 6 b       A             1  
 7 a       A             2  
 8 b       A             2  
 9 a       B             0.5
10 b       B             0.5
11 a       B             1  
12 b       B             1  
13 a       B             2  
14 b       B             2  
\end{verbatim}

\subsection{Links}\label{links}

In {edibble}, each experimental factor is encoded as a node in the
factor graph along with its levels as nodes in the level graph
(Figure~\ref{fig-graphs}). The edges (or links) can only be specified
after the nodes are created. The links define the relationship between
the experimental factors and the direction determining the hierarchy
with the nodes (see Tanaka 2023 for more information). Often, these
links are implicitly understood and not explicitly encoded, thus making
it difficult to utilise the information downstream. By encoding the
links, we can derive information and validate processes downstream.

Users specify these links using functions that are semantically aligned
with thinking in the construction of an experimental design. There are
three high-level approaches to defining these links as summarised in
Table~\ref{tbl-links} and explained in more detail next.

\begin{longtable}[]{@{}
  >{\raggedright\arraybackslash}p{(\columnwidth - 6\tabcolsep) * \real{0.2000}}
  >{\raggedright\arraybackslash}p{(\columnwidth - 6\tabcolsep) * \real{0.2000}}
  >{\raggedright\arraybackslash}p{(\columnwidth - 6\tabcolsep) * \real{0.2000}}
  >{\raggedright\arraybackslash}p{(\columnwidth - 6\tabcolsep) * \real{0.4000}}@{}}
\caption{The table lists the three main approaches to specify links
between experimental factors or their levels, the functions relevant for
each approach, which graph the functions modifies, and the general
purpose of the functions.}\label{tbl-links}\tabularnewline
\toprule\noalign{}
\begin{minipage}[b]{\linewidth}\raggedright
Approach
\end{minipage} & \begin{minipage}[b]{\linewidth}\raggedright
Functions
\end{minipage} & \begin{minipage}[b]{\linewidth}\raggedright
Modifies
\end{minipage} & \begin{minipage}[b]{\linewidth}\raggedright
Purpose
\end{minipage} \\
\midrule\noalign{}
\endfirsthead
\toprule\noalign{}
\begin{minipage}[b]{\linewidth}\raggedright
Approach
\end{minipage} & \begin{minipage}[b]{\linewidth}\raggedright
Functions
\end{minipage} & \begin{minipage}[b]{\linewidth}\raggedright
Modifies
\end{minipage} & \begin{minipage}[b]{\linewidth}\raggedright
Purpose
\end{minipage} \\
\midrule\noalign{}
\endhead
\bottomrule\noalign{}
\endlastfoot
Within role group & {nested\_in}, {crossed\_by}, {conditioned\_on} &
Both factor and level graphs & Links between the nodes of the same role
only. \\
Allotment & {allot\_trts}, {allot\_units}, {set\_rcrds},
{set\_rcrds\_of} & Factor graph only & Capture high-level links that are
typically apriori known by the user. \\
Assignment & {assign\_trts}, {assign\_units} & Level graph only &
Determine links between nodes, often algorithmically. \\
\end{longtable}

\subsubsection{Within role group}\label{sec-within}

The helper functions, {nested\_in} and {crossed\_by}, are demonstrated
in Section~\ref{sec-nested} and Section~\ref{sec-crossed}, to construct
nested and crossed units, respectively. The helper function,
{conditioned\_on}, demonstrated in Section~\ref{sec-conditioned},
constructs a conditional treatment structure. These helper functions
concurrently draw links between the relevant nodes in both factor and
level graphs. These links would be apriori known to the user and these
helper functions are just semantically designed to make it easier for
the user to specify the links between nodes. These helper functions only
construct links between nodes belonging to the same role (i.e., unit or
treatment).

\subsubsection{Allotment}\label{sec-allot}

Links specified using an allotment approach designate high-level links
between factors. In other words, this approach only draws edges between
nodes in the factor graph, and almost always, these edges are
intentionally formed by the user. The purpose of this approach is to
capture a user's high-level intention or knowledge.

For demonstration, we leverage the previously defined unit
(\texttt{complexd}) and treatment structures (\texttt{factrtc}). These
structures can be combined to obtain the combined design object as
below.

\begin{Shaded}
\begin{Highlighting}[]
\NormalTok{complexd }\SpecialCharTok{+}\NormalTok{ factrtc}
\end{Highlighting}
\end{Shaded}

\begin{verbatim}
Complex structure with different dimensions
+-site (4 levels)
| +-col (30 levels)
| | \-plot (90 levels)
| +-row (12 levels)
| | \-plot (90 levels)
| \-plot (90 levels)
+-variety (2 levels)
+-fertilizer (3 levels)
\-amount (4 levels)
\end{verbatim}

The above design object does not describe the links between the
treatments and units. The function {allot\_trts} ascribes the links
between treatments to units in the factor graph.

\begin{Shaded}
\begin{Highlighting}[]
\NormalTok{alloted1 }\OtherTok{\textless{}{-}}\NormalTok{ (complexd }\SpecialCharTok{+}\NormalTok{ factrtc) }\SpecialCharTok{\%\textgreater{}\%} 
  \FunctionTok{allot\_trts}\NormalTok{(    fertilizer }\SpecialCharTok{\textasciitilde{}}\NormalTok{ row,}
\NormalTok{             amount}\SpecialCharTok{:}\NormalTok{variety }\SpecialCharTok{\textasciitilde{}}\NormalTok{ plot)}
\end{Highlighting}
\end{Shaded}

\subsubsection{Assignment}\label{sec-assign}

\begin{figure}

\centering{

\includegraphics{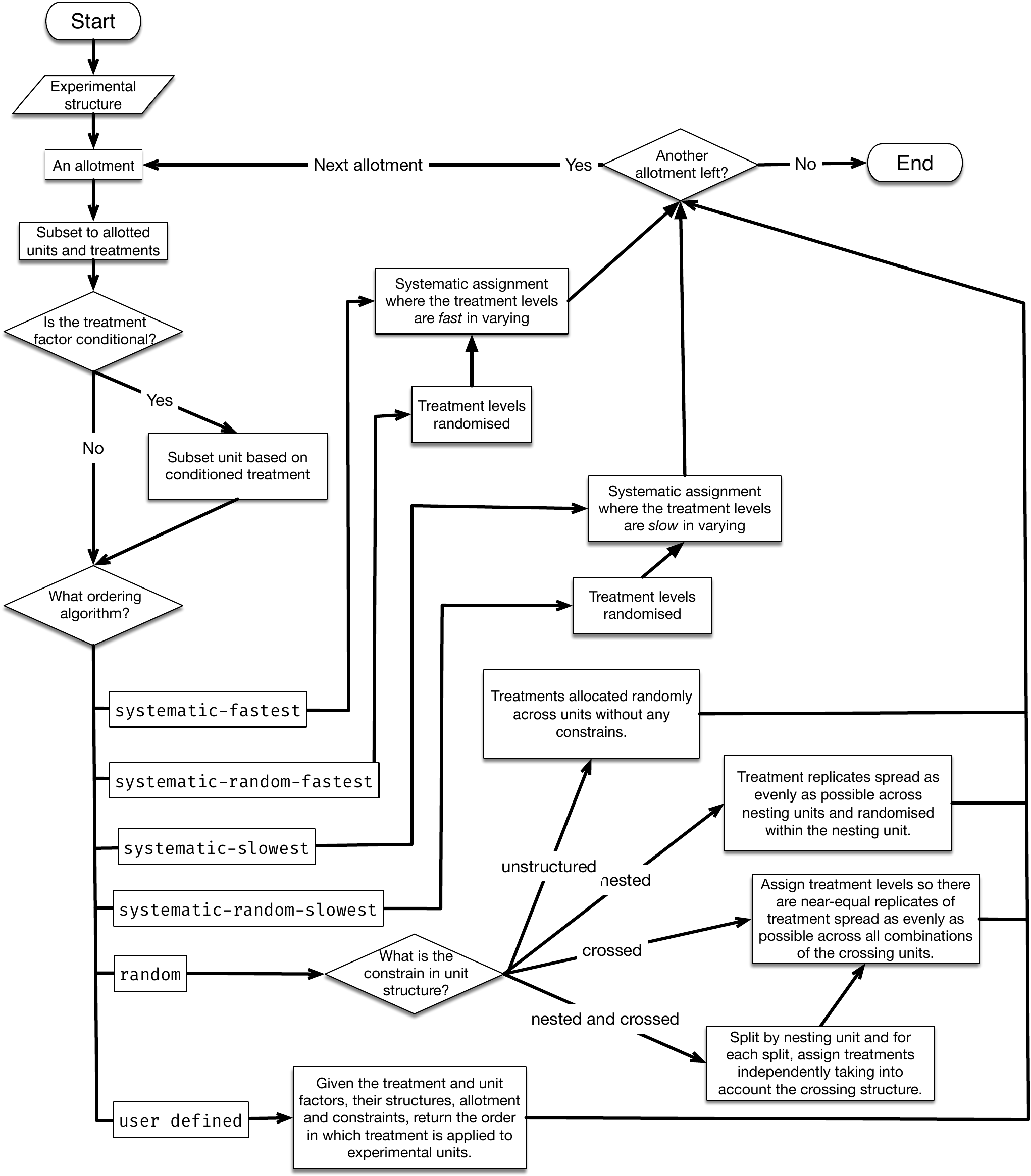}

}

\caption{\label{fig-assign-alg}An overview of the treatment assignment
algorithm in {assign\_trts}. For the given experimental structure, the
algorithm iterates over the defined allotments. Each treatment to unit
allotment then generates an order of treatment levels assigned to units
based on the ordering algorithm selected.}

\end{figure}

The {assign\_trts} (often algorithmically) draw links between the
treatment and unit nodes in the level graph (conditioned on the existing
links in the factor graph). An overview of the algorithm in
{assign\_trts} is shown in Figure~\ref{fig-assign-alg}.

\begin{figure}

\centering{

\includegraphics{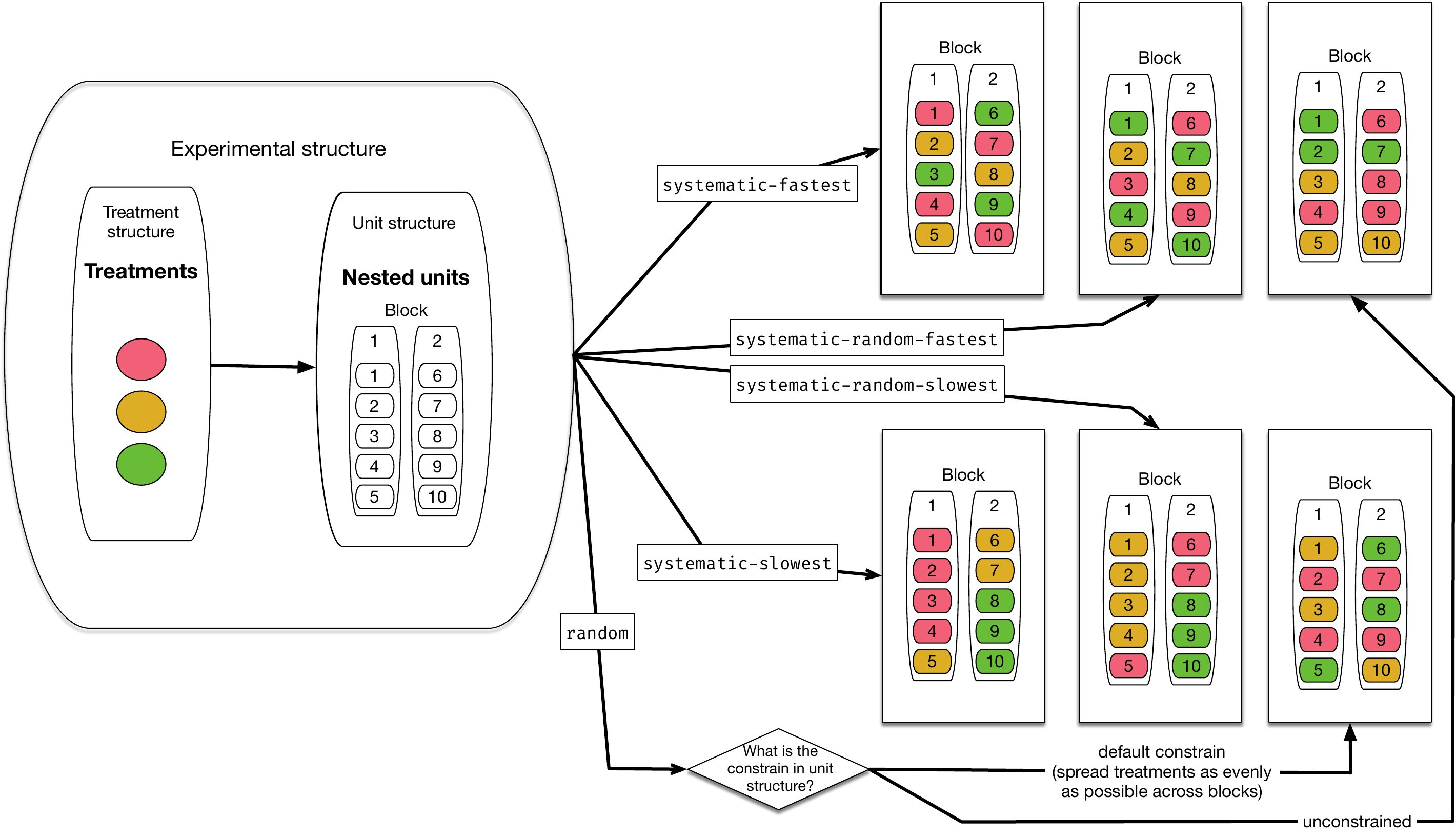}

}

\caption{\label{fig-actual-assign}The left shows the treatment structure
(3 treatment levels) and unit structure (2 blocks with 5 units nested in
each, respresented by the rounded rectangle and numbers within it
representing the unit order) that collectively make up the experimental
structure. The treatment is alloted to unit factor. The actual
assignment depends on the ordering algorithm selected with six
variations shown on the right.}

\end{figure}

There are five in-built assignment algorithms: ``systematic-fastest''
(synonym for ``systematic''), ``systematic-random-fastest'' (synonym for
``systematic-random''), ``systematic-slowest'',
``systematic-random-slowest'', and ``random''. The variation in
systematic assignment results in repeated ordering with respect to the
unit order, without regard to any unit structure. When the number of
units is not divisible by the total number of treatments, the earlier
treatment levels would have an extra replicate. The
``systematic-random-fastest'' and ``systematic-random-slowest'' are
systematic variants that ensure equal chances for all treatment levels
to obtain an extra replicate by randomising the order of treatment
levels before the systematic allocation of treatment to units proceeds.
The ``fastest'' and ``slowest'' variants determine if treatment levels
are fast or slow in varying across order of the unit (slow varying
meaning that the same treatment levels will be closer together in unit
order, whereas fast varying means the same treatment levels are spread
out in unit order). An example, with three treatments allotted to ten
units, illustrating the in-built assignment algorithms is shown in
Figure~\ref{fig-actual-assign}.

\begin{figure}

\centering{

\includegraphics{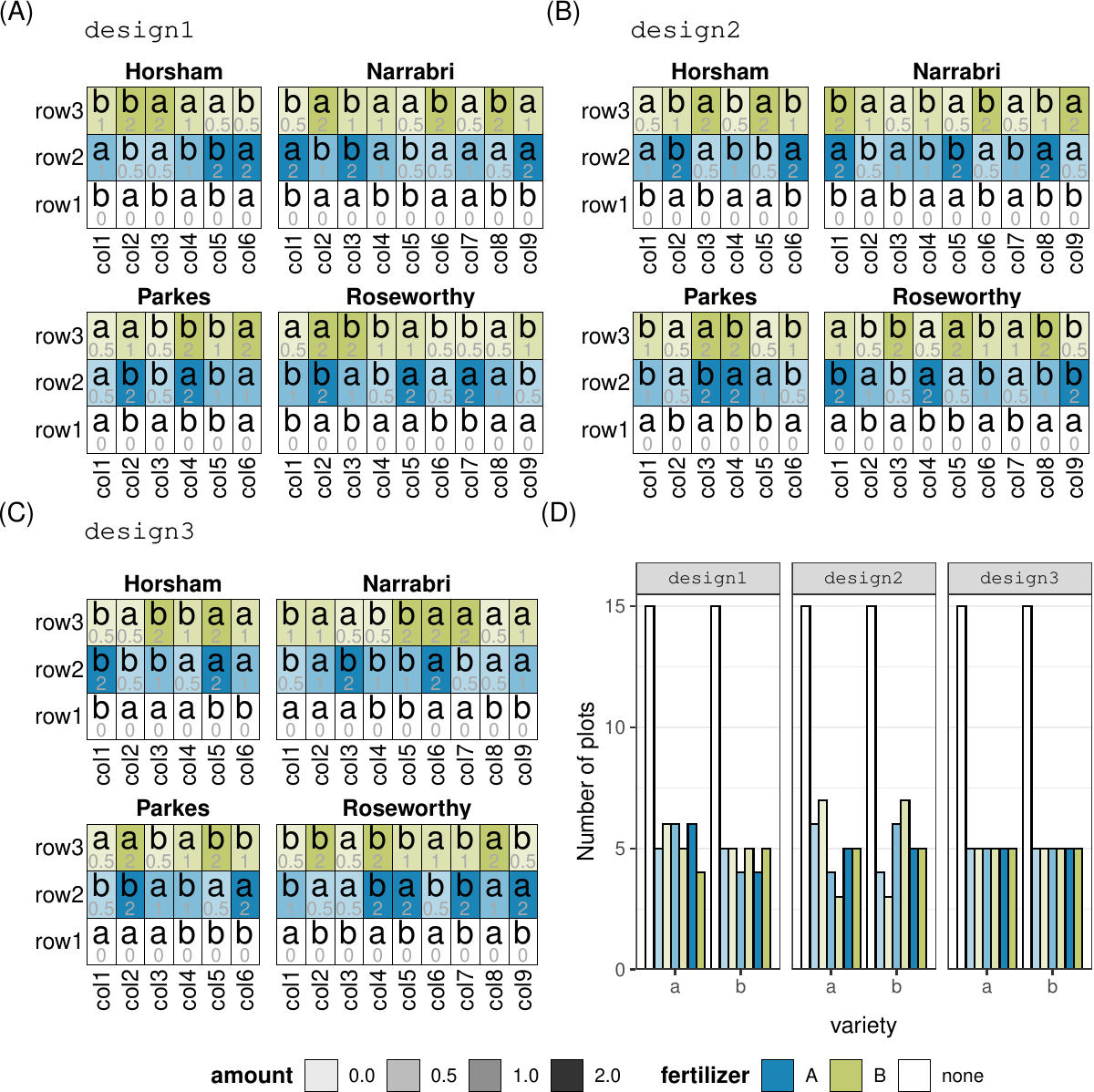}

}

\caption{\label{fig-design-output}(A)-(C) shows the treatment allocation
to plots from the three generated designs (\texttt{design1},
\texttt{design2}, \texttt{design3}). A plot is represented as a tile
laid out in a rectangular array for each site. The color of the tile
shows the type of fertilizer that the plot was allocated. The opacity of
the color represents the amount of fertilizer allocated to the plot (the
amount is also written in each tile in grey). The variety (a or b)
allocated to the plot is written on each tile. (D) shows the number of
plots that were alloted particular level of the treatment for each of
the three designs.}

\end{figure}

Building on the previously defined structure and allotment, we define an
algorithm to assign links between unit and treatment levels using the
function {assign\_trts}. Below, we use a systematic ordering for the
first allotment (fertilizer to row) then a random ordering for the
second allotment (interaction of amount and variety to plot). An
optional seed number is provided to ensure the generated design could be
reproduced. The generated design is shown in
Figure~\ref{fig-design-output} (A).

\begin{Shaded}
\begin{Highlighting}[]
\NormalTok{design1 }\OtherTok{\textless{}{-}}\NormalTok{ alloted1 }\SpecialCharTok{\%\textgreater{}\%} 
  \FunctionTok{assign\_trts}\NormalTok{(}\AttributeTok{order =} \FunctionTok{c}\NormalTok{(}\StringTok{"systematic"}\NormalTok{, }\StringTok{"random"}\NormalTok{),}
              \AttributeTok{seed =} \DecValTok{2023}\NormalTok{) }\SpecialCharTok{\%\textgreater{}\%} 
  \FunctionTok{serve\_table}\NormalTok{(}\AttributeTok{label\_nested =} \FunctionTok{c}\NormalTok{(row, col))}
\end{Highlighting}
\end{Shaded}

While allotment (high-level allocation) and assignment (actual
allocation) are distinguished in the system to provide flexibility to
the user for defining these processes separately, it is likely that many
users would concurrently define these processes. The {allot\_table}
function offers a shorthand that combines the call to {allot\_trts},
{assign\_trts}, and {serve\_table} into one call.

To illustrate the difference when treatment interaction is alloted to a
unit (like the second allotment in \texttt{allotment1}), below, we have
a different allotment where the amount of fertilizer and variety are
allotted to plot in a separate allotment. A separate allotment can be
assigned using different algorithms and is considered independent of
other allotments (unless the treatment factor is conditional on another
treatment factor). The resulting assignment is shown in
Figure~\ref{fig-design-output} (B) for the design below.

\begin{Shaded}
\begin{Highlighting}[]
\NormalTok{design2 }\OtherTok{\textless{}{-}}\NormalTok{ (complexd }\SpecialCharTok{+}\NormalTok{ factrtc) }\SpecialCharTok{\%\textgreater{}\%} 
  \FunctionTok{allot\_table}\NormalTok{(fertilizer }\SpecialCharTok{\textasciitilde{}}\NormalTok{ row,}
\NormalTok{                  amount }\SpecialCharTok{\textasciitilde{}}\NormalTok{ plot,}
\NormalTok{                 variety }\SpecialCharTok{\textasciitilde{}}\NormalTok{ plot, }
              \AttributeTok{order =} \FunctionTok{c}\NormalTok{(}\StringTok{"systematic"}\NormalTok{, }\StringTok{"random"}\NormalTok{, }\StringTok{"random"}\NormalTok{),}
              \AttributeTok{label\_nested =} \FunctionTok{c}\NormalTok{(row, col),}
              \AttributeTok{seed =} \DecValTok{2023}\NormalTok{)}
\end{Highlighting}
\end{Shaded}

The assignment algorithms in the system use the default constraint,
which takes the nesting structure defined in the unit structure
(i.e.~row is nested in site and plot is crossed by row and column and
nested in site). This constraint is used to define the nature of
``random'' assignment. For example, in the code below, we relax this
constraint such that the \texttt{plot} factor is constrained within a
\texttt{row} (default was \texttt{row}, \texttt{col} and \texttt{site}),
which in turn is contained within the \texttt{site}. This difference in
constraints results in a different path in the algorithm (as shown in
the overview in Figure~\ref{fig-assign-alg}), which generates the design
shown in Figure~\ref{fig-design-output} (C).

\begin{Shaded}
\begin{Highlighting}[]
\NormalTok{design3 }\OtherTok{\textless{}{-}}\NormalTok{ alloted1 }\SpecialCharTok{\%\textgreater{}\%} 
  \FunctionTok{assign\_trts}\NormalTok{(}\AttributeTok{order =} \FunctionTok{c}\NormalTok{(}\StringTok{"systematic"}\NormalTok{, }\StringTok{"random"}\NormalTok{),}
              \AttributeTok{seed =} \DecValTok{2023}\NormalTok{, }
              \AttributeTok{constrain =} \FunctionTok{list}\NormalTok{(}\AttributeTok{row =} \StringTok{"site"}\NormalTok{, }\AttributeTok{plot =} \StringTok{"row"}\NormalTok{)) }\SpecialCharTok{\%\textgreater{}\%} 
  \FunctionTok{serve\_table}\NormalTok{(}\AttributeTok{label\_nested =} \FunctionTok{c}\NormalTok{(row, col))}
\end{Highlighting}
\end{Shaded}

The above three different designs (\texttt{design1}, \texttt{design2}
and \texttt{design3}) share the same unit and treatment structure, but
the allotment and/or assignment algorithm differed. One result of this
is that the treatment replications, as shown in
Figure~\ref{fig-design-output} (D), differ across the generated designs
with the most ideal distribution seen in \texttt{design3} (if all
fertilizer and amount combinations are of equal interest and fertilizer
allocation is restricted to the row; arguably, it is better to remove
the latter constraint, if practically feasible, so the units with the
control treatment can be assigned for other treatment levels to obtain a
more even distribution). The difference in \texttt{design1} and
\texttt{design2} is that the amount and variety were allocated as an
interaction in the former but independently in the latter. The latter
process does not ensure near-equal replication of the treatment levels,
so it is not surprising that in Figure~\ref{fig-design-output} (D),
\texttt{design2} has the least uniform treatment distribution.

Finding or creating the most appropriate assignment algorithm is one of
the challenging tasks in the whole workflow. The default algorithm is
unlikely to be optimal for the given structure, and the user is
encouraged to modify this step to suit their own design.
Section~\ref{sec-composition} presents an example of this.

\subsection{Records}\label{sec-rcrds}

The values of a record factor is unknown prior to the execution of the
experiment; thus, the method used to define it fundamentally differs
from other factors. The record factor is defined using {set\_rcrds} with
the arguments in the form of \texttt{record\ =\ unit}, where
\texttt{record} corresponds to user-specified name of the record, and
\texttt{unit} corresponds to an existing unit factor. The LHS of the
input argument is always a new factor, as it was the case for
{set\_units} and {set\_trts}. Multiple record factors can be specified
as separate arguments within the same function call as shown below.

\begin{Shaded}
\begin{Highlighting}[]
\NormalTok{record1 }\OtherTok{\textless{}{-}}\NormalTok{ design1 }\SpecialCharTok{\%\textgreater{}\%} 
  \FunctionTok{set\_rcrds}\NormalTok{(}\AttributeTok{biomass =}\NormalTok{ plot,}
              \AttributeTok{yield =}\NormalTok{ plot, }
           \AttributeTok{rainfall =}\NormalTok{ site)}
\end{Highlighting}
\end{Shaded}

The above specification can be awkwardly lengthy when we expect
multivariate data (i.e., multiple responses of the same unit). In this
instance, it may be easier to specify the responses as follows.

\begin{Shaded}
\begin{Highlighting}[]
\NormalTok{design1 }\SpecialCharTok{\%\textgreater{}\%} 
  \FunctionTok{set\_rcrds\_of}\NormalTok{(}\AttributeTok{plot =} \FunctionTok{c}\NormalTok{(}\StringTok{"biomass"}\NormalTok{, }\StringTok{"yield"}\NormalTok{),}
               \AttributeTok{site =} \StringTok{"rainfall"}\NormalTok{)}
\end{Highlighting}
\end{Shaded}

In the above specification, the LHS corresponds to an existing unit
factor, while the RHS is a character vector where each element
corresponds to the name of the record factor. Notice that in this
specification, the RHS elements are quoted as they do not yet exist in
the design. By contrast, the RHS specification in {set\_rcrds} is
unquoted because these factors already exist. Because the intention of
the LHS specification is to point to an existing unit factor, it differs
from the convention of other functions that prefix with \texttt{set\_}.
As a signal to the different input structure, the function appends a
suffix \texttt{\_of}, which is supposed to act as a signal to the user
that this is different from other \texttt{set\_} functions.

Record factors generally do not play a role in the assignment of
treatments to units; however, they play a critical role in the analysis
stage. The specification of these record factors allows for the user's
intention to be explicit. The added benefits also include the encoding
of the expected and simulated values, as described in detail next.

\subsubsection{Expected values}\label{sec-expect}

If the record factors are defined in the design object, like in
\texttt{record1}, then the user may specify the expected values of the
records. For example, below we specify that \texttt{biomass} is greater
than or equal to zero, yield is between 0 and 10, and rainfall is a
factor with two levels: high or low.

\begin{Shaded}
\begin{Highlighting}[]
\NormalTok{expect1 }\OtherTok{\textless{}{-}}\NormalTok{ record1 }\SpecialCharTok{\%\textgreater{}\%} 
  \FunctionTok{expect\_rcrds}\NormalTok{(biomass }\SpecialCharTok{\textgreater{}=} \DecValTok{0}\NormalTok{,}
\NormalTok{               yield }\SpecialCharTok{\textgreater{}} \DecValTok{0}\NormalTok{, yield }\SpecialCharTok{\textless{}} \DecValTok{10}\NormalTok{,}
               \FunctionTok{factor}\NormalTok{(rainfall, }\AttributeTok{levels =} \FunctionTok{c}\NormalTok{(}\StringTok{"high"}\NormalTok{, }\StringTok{"low"}\NormalTok{)))}
\end{Highlighting}
\end{Shaded}

Once the expected values are encoded, they can be used in various
functions. For example, suppose we export this design as below.

\begin{Shaded}
\begin{Highlighting}[]
\FunctionTok{export\_design}\NormalTok{(expect1, }\AttributeTok{file =} \StringTok{"mydesign.xlsx"}\NormalTok{, }\AttributeTok{overwrite =} \ConstantTok{TRUE}\NormalTok{)}
\end{Highlighting}
\end{Shaded}

\begin{figure}

\centering{

\includegraphics{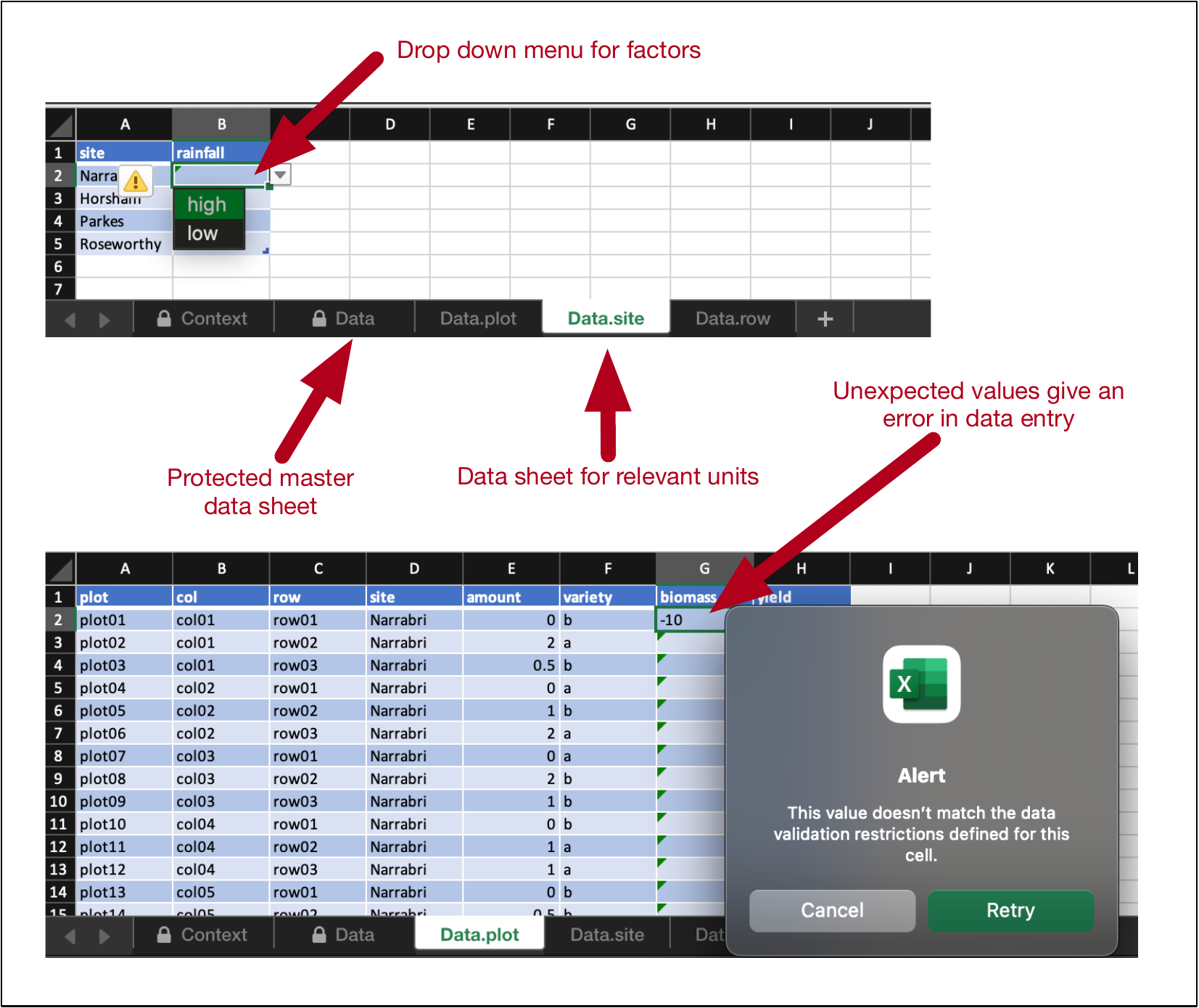}

}

\caption{\label{fig-export}The exported design in the Excel sheet
encodes the data validation defined in the design object such that the
unexpected values cannot be entered in the corresponding entry. The data
entry is designed to remove redundancy of duplicate entries (e.g.~the
rainfall status recorded for site, but yield and biomass on the plot) by
creating separate data sheets for the corresponding unit factor.}

\end{figure}

The exported design automatically embeds the data validation in the
corresponding cells, as shown in Figure~\ref{fig-export}.

\subsubsection{Simulated values}\label{sec-simulate}

Another benefit of encoding the expected values of the record factors is
that you can do a lazy simulation using {autofill\_rcrds}. This function
randomly chooses a simulation scheme (including the variables that
influence the record), while ensuring that it keeps to the expected
values.

\begin{Shaded}
\begin{Highlighting}[]
\FunctionTok{set.seed}\NormalTok{(}\DecValTok{2023}\NormalTok{)}
\NormalTok{sim1auto }\OtherTok{\textless{}{-}} \FunctionTok{autofill\_rcrds}\NormalTok{(expect1) }
\end{Highlighting}
\end{Shaded}

The aforementioned lazy simulation is designed for quick diagnostics of
planned analytical methods and not for any serious simulation study. For
proper simulation schemes, users can freely enter their own schemes
using the function {simulate\_process}. In this function, the user
defines a series of functions, where each function returns either 1) a
vector of the same size as the data or 2) a {data.frame} of the same row
dimension as the data. For 1) the name of the function must correspond
to the name of the record factor, and for 2) the name must start with a
dot and the column names of the returning object must match the names of
the record factors.

In the example below, we define two simulation processes called
``yield'' and ``.multi''. The first simulation process is a function
that has arguments that define the variety main effect and combined
effect of fertilizer and its amount; the return object is a numeric
vector where random noise is added on top of the fixed treatment
effects. The second simulation process is a function that simulates
yield and biomass from a multivariate normal distribution.

\begin{Shaded}
\begin{Highlighting}[]
\NormalTok{process1 }\OtherTok{\textless{}{-}}\NormalTok{ expect1 }\SpecialCharTok{\%\textgreater{}\%} 
  \FunctionTok{simulate\_process}\NormalTok{(}
    \AttributeTok{yield =} \ControlFlowTok{function}\NormalTok{(}\AttributeTok{v =} \FunctionTok{c}\NormalTok{(}\DecValTok{2}\NormalTok{, }\DecValTok{4}\NormalTok{),}
                     \AttributeTok{af =} \FunctionTok{c}\NormalTok{(}\DecValTok{1}\NormalTok{, }\FloatTok{2.5}\NormalTok{, }\DecValTok{3}\NormalTok{, }\FloatTok{0.5}\NormalTok{, }\DecValTok{1}\NormalTok{, }\DecValTok{5}\NormalTok{, }\DecValTok{0}\NormalTok{)) \{}
\NormalTok{      veffects }\OtherTok{\textless{}{-}} \FunctionTok{setNames}\NormalTok{(v, }\FunctionTok{c}\NormalTok{(}\StringTok{"a"}\NormalTok{, }\StringTok{"b"}\NormalTok{))}
\NormalTok{      afeffects }\OtherTok{\textless{}{-}} \FunctionTok{setNames}\NormalTok{(af, }\FunctionTok{c}\NormalTok{(}\StringTok{"A:0.5"}\NormalTok{, }\StringTok{"A:1"}\NormalTok{, }\StringTok{"A:2"}\NormalTok{,}
                                  \StringTok{"B:0.5"}\NormalTok{, }\StringTok{"B:1"}\NormalTok{, }\StringTok{"B:2"}\NormalTok{, }\StringTok{"none:0"}\NormalTok{))}
      \CommentTok{\# must return vector of the same size at the data}
\NormalTok{      veffects[variety] }\SpecialCharTok{+}\NormalTok{ afeffects[}\FunctionTok{paste0}\NormalTok{(fertilizer, }\StringTok{":"}\NormalTok{, amount)] }\SpecialCharTok{+} \FunctionTok{rnorm}\NormalTok{(}\FunctionTok{n}\NormalTok{())}
\NormalTok{    \},}
    
    \AttributeTok{.multi =} \ControlFlowTok{function}\NormalTok{() \{}
\NormalTok{      Sigma }\OtherTok{\textless{}{-}} \FunctionTok{matrix}\NormalTok{(}\FunctionTok{c}\NormalTok{(}\DecValTok{2}\NormalTok{, }\FloatTok{0.9}\NormalTok{, }\FloatTok{0.9}\NormalTok{, }\DecValTok{1}\NormalTok{), }\AttributeTok{nrow =} \DecValTok{2}\NormalTok{)}
\NormalTok{      yveffects }\OtherTok{\textless{}{-}} \FunctionTok{c}\NormalTok{(}\StringTok{"a"} \OtherTok{=} \DecValTok{3}\NormalTok{, }\StringTok{"b"} \OtherTok{=} \DecValTok{9}\NormalTok{)}
\NormalTok{      bveffects }\OtherTok{\textless{}{-}} \FunctionTok{c}\NormalTok{(}\StringTok{"a"} \OtherTok{=} \DecValTok{1}\NormalTok{, }\StringTok{"b"} \OtherTok{=} \DecValTok{3}\NormalTok{)}
\NormalTok{      res }\OtherTok{\textless{}{-}} \FunctionTok{cbind}\NormalTok{(yveffects[variety], bveffects[variety])}
\NormalTok{      res }\OtherTok{\textless{}{-}}\NormalTok{ res }\SpecialCharTok{+}\NormalTok{ mvtnorm}\SpecialCharTok{::}\FunctionTok{rmvnorm}\NormalTok{(}\FunctionTok{n}\NormalTok{(), }\AttributeTok{mean =} \FunctionTok{c}\NormalTok{(}\DecValTok{0}\NormalTok{, }\DecValTok{0}\NormalTok{), }\AttributeTok{sigma =}\NormalTok{ Sigma)}
\NormalTok{      res }\OtherTok{\textless{}{-}} \FunctionTok{as.data.frame}\NormalTok{(res)}
      \FunctionTok{colnames}\NormalTok{(res) }\OtherTok{\textless{}{-}} \FunctionTok{c}\NormalTok{(}\StringTok{"yield"}\NormalTok{, }\StringTok{"biomass"}\NormalTok{)}
      \CommentTok{\# must return a data.frame where the column names match the}
      \CommentTok{\# record factor names and the number of rows match the }
      \CommentTok{\# row dimension of data}
\NormalTok{      res}
\NormalTok{    \}}
\NormalTok{  )}
\end{Highlighting}
\end{Shaded}

After defining the simulation process, the actual simulation of the
record is performed through a call to {simulate\_rcrds}. Below, we
simulate the actual records by naming the processes we want to use along
with any parameters of the function. Optionally, users can define values
to censor the simulated values if they lie outside the expected values.

\begin{Shaded}
\begin{Highlighting}[]
\NormalTok{sim1default }\OtherTok{\textless{}{-}} \FunctionTok{simulate\_rcrds}\NormalTok{(process1, }\AttributeTok{yield =} \FunctionTok{with\_params}\NormalTok{())}
\NormalTok{sim1var }\OtherTok{\textless{}{-}} \FunctionTok{simulate\_rcrds}\NormalTok{(process1, }
                          \AttributeTok{yield =} \FunctionTok{with\_params}\NormalTok{(}\AttributeTok{v =} \FunctionTok{c}\NormalTok{(}\SpecialCharTok{{-}}\DecValTok{1}\NormalTok{, }\DecValTok{5}\NormalTok{), }
                                              \AttributeTok{.censor =} \FunctionTok{c}\NormalTok{(}\DecValTok{0}\NormalTok{, }\DecValTok{10}\NormalTok{)))}
\NormalTok{sim1multi }\OtherTok{\textless{}{-}} \FunctionTok{simulate\_rcrds}\NormalTok{(process1, }\AttributeTok{.multi =} \FunctionTok{with\_params}\NormalTok{(), }\AttributeTok{.seed =} \DecValTok{1}\NormalTok{)}
\NormalTok{sim1censor }\OtherTok{\textless{}{-}} \FunctionTok{simulate\_rcrds}\NormalTok{(process1, }
                \AttributeTok{.multi =} \FunctionTok{with\_params}\NormalTok{(}\AttributeTok{.censor =} \FunctionTok{list}\NormalTok{(}\AttributeTok{yield =} \FunctionTok{c}\NormalTok{(}\DecValTok{0}\NormalTok{, }\DecValTok{10}\NormalTok{),}
                                                    \AttributeTok{biomass =} \DecValTok{0}\NormalTok{)),}
                \AttributeTok{.seed =} \DecValTok{1}\NormalTok{)}
\end{Highlighting}
\end{Shaded}

\begin{figure}

\centering{

\includegraphics{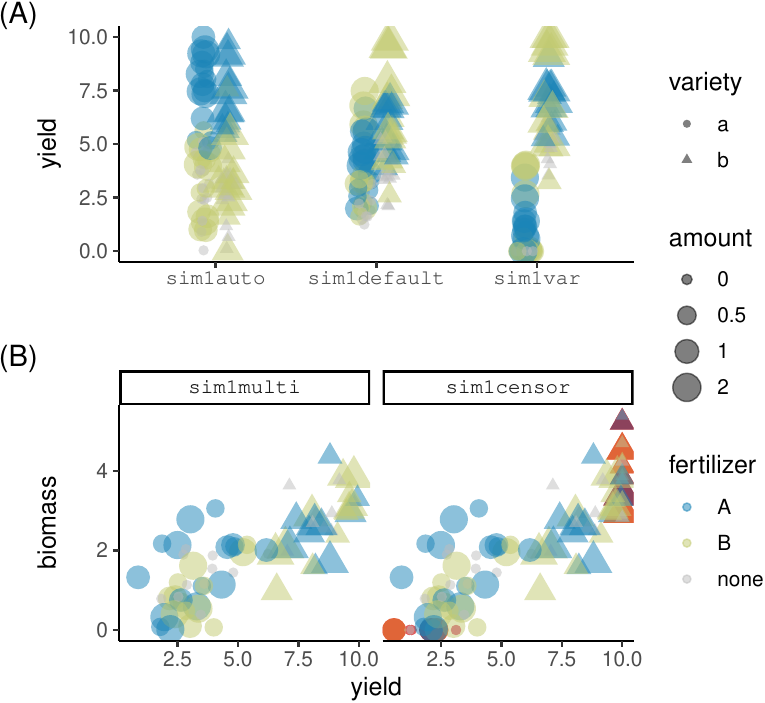}

}

\caption{\label{fig-simulation}(A) show the simulated yields from
\texttt{sim1auto}, \texttt{sim1default} and \texttt{sim1var} colored by
fertilizer type with shape showing the variety, and size reflecting the
amount of fertilizer. (B) show the scatterplot of biomass and yield from
\texttt{sim1multi} and \texttt{sim1censor}. The only difference in the
two simulated values are the censored values that are shown as red
points in \texttt{sim1censor}.}

\end{figure}

The resulting simulation values from the above calls are shown in
Figure~\ref{fig-simulation}. We can see in Figure~\ref{fig-simulation}
(A) that the \texttt{sim1auto} shows a huge difference in yield with
respect to fertilizer type (reflecting that the automated simulation
process must have included fertilizer type in the data generation
process); \texttt{sim1default} shows some difference in yield across
variety (the actual difference is 2 in the simulation process); and
\texttt{sim1var} shows a variation in the latter process where the
difference in the main variety effects are now more pronounced (actual
difference is 6). Figure~\ref{fig-simulation} (B) shows the simulation
results from \texttt{sim1multi} and \texttt{sim1censor}; the difference
in the latter is that the values are censored to 0 or 10 in the yield
and 0 in biomass (the extra points in yield at 0 and 10, and biomass at
0 are easily seen in the plot).

\section{Worked examples}\label{sec-examples}

In addition to the example presented in Section~\ref{sec-usage}, we
demonstrate two examples from real experiments: one in ecology and one
in linguistics.

\subsection{Common garden design}\label{common-garden-design}

Cochrane et al. (2015) studied the temperature and moisture impact on
seedling emergence in a common garden. Below is a digest of the
experimental design description.

\begin{tcolorbox}[enhanced jigsaw, breakable, opacitybacktitle=0.6, leftrule=.75mm, colback=white, opacityback=0, colbacktitle=quarto-callout-note-color!10!white, title=\textcolor{quarto-callout-note-color}{\faInfo}\hspace{0.5em}{Example 2: Common garden design}, bottomrule=.15mm, arc=.35mm, coltitle=black, bottomtitle=1mm, toprule=.15mm, toptitle=1mm, left=2mm, titlerule=0mm, colframe=quarto-callout-note-color-frame, rightrule=.15mm]

Twelve shelters were constructed on site to manipulate water
availability and assess the effects on seed germination, seedling
emergence, and seedling growth and survival. Two raised garden beds were
placed below each shelter. Water was collected from the roof of each
shelter and passively irrigated to the garden beds in each of the three
water treatments: reduced rainfall R (80\% of the total), natural
rainfall N (100\% of the total), and increased rainfall I (140\% of the
total). Each shelter contained one garden bed fitted with 24 warming
chambers (W), and one garden bed without warming chambers (C).

Four non-sprouting, obligate seeding species (\emph{Banksia media},
\emph{Banksia coccinea}, \emph{Banksia baxteri}, and \emph{Banksia
quercifolia}) were chosen for the study. Six discrete populations of
each species were selected to represent a rainfall cline (High, Medium
or Low). The term `population' is used to describe plants originating
from a particular geographic and climatic location.

The experimental units were arranged within each garden bed in a
balanced array of three columns and eight rows, with seeds from all 24
populations sown in each garden bed. Each row contained seeds from a
single species. Each column in the array contained seeds from eight
populations corresponding to a particular level of rainfall. Within
these constraints, the populations were divided into two sets of 12, so
that the same 12 populations always occurred together in a set of four
rows.

The allocation of seeds to the experimental units was fully randomised,
subject to the constraints of the design.

\end{tcolorbox}

Based on the above description, my understanding of the unit structure
was follows.

\begin{Shaded}
\begin{Highlighting}[]
\NormalTok{garden\_units1 }\OtherTok{\textless{}{-}} \FunctionTok{design}\NormalTok{(}\StringTok{"Common garden experiment"}\NormalTok{) }\SpecialCharTok{\%\textgreater{}\%} 
  \FunctionTok{set\_units}\NormalTok{(}\AttributeTok{shelter =} \DecValTok{12}\NormalTok{,}
            \AttributeTok{bed =} \FunctionTok{nested\_in}\NormalTok{(shelter, }\DecValTok{2}\NormalTok{),}
            \AttributeTok{row =} \FunctionTok{nested\_in}\NormalTok{(bed, }\DecValTok{8}\NormalTok{),}
            \AttributeTok{col =} \FunctionTok{nested\_in}\NormalTok{(bed, }\DecValTok{3}\NormalTok{),}
            \AttributeTok{plot =} \FunctionTok{nested\_in}\NormalTok{(bed, }\FunctionTok{crossed\_by}\NormalTok{(row, col)))}
\end{Highlighting}
\end{Shaded}

Initially, I did not understand the full treatment structure and wrote
only partial treatment factors. From a further dissection of the
description, I inferred that rainfall was the other treatment condition.
At this point it was unclear to me how their so-called ``population'' is
actually mapped to the rainfall and so I ignore it for the moment.

\begin{Shaded}
\begin{Highlighting}[]
\NormalTok{garden\_trts1 }\OtherTok{\textless{}{-}} \FunctionTok{set\_trts}\NormalTok{(}\AttributeTok{water =} \FunctionTok{c}\NormalTok{(}\StringTok{"R"}\NormalTok{, }\StringTok{"N"}\NormalTok{, }\StringTok{"I"}\NormalTok{),}
                         \AttributeTok{chamber =} \FunctionTok{c}\NormalTok{(}\StringTok{"W"}\NormalTok{, }\StringTok{"C"}\NormalTok{), }
                         \AttributeTok{species =} \FunctionTok{c}\NormalTok{(}\StringTok{"media"}\NormalTok{, }\StringTok{"coccinea"}\NormalTok{,}
                                     \StringTok{"baxteri"}\NormalTok{, }\StringTok{"quercifolia"}\NormalTok{),}
                         \AttributeTok{rainfall =} \FunctionTok{c}\NormalTok{(}\StringTok{"High"}\NormalTok{, }\StringTok{"Medium"}\NormalTok{, }\StringTok{"Low"}\NormalTok{))}
\end{Highlighting}
\end{Shaded}

\begin{figure}

\centering{

\includegraphics{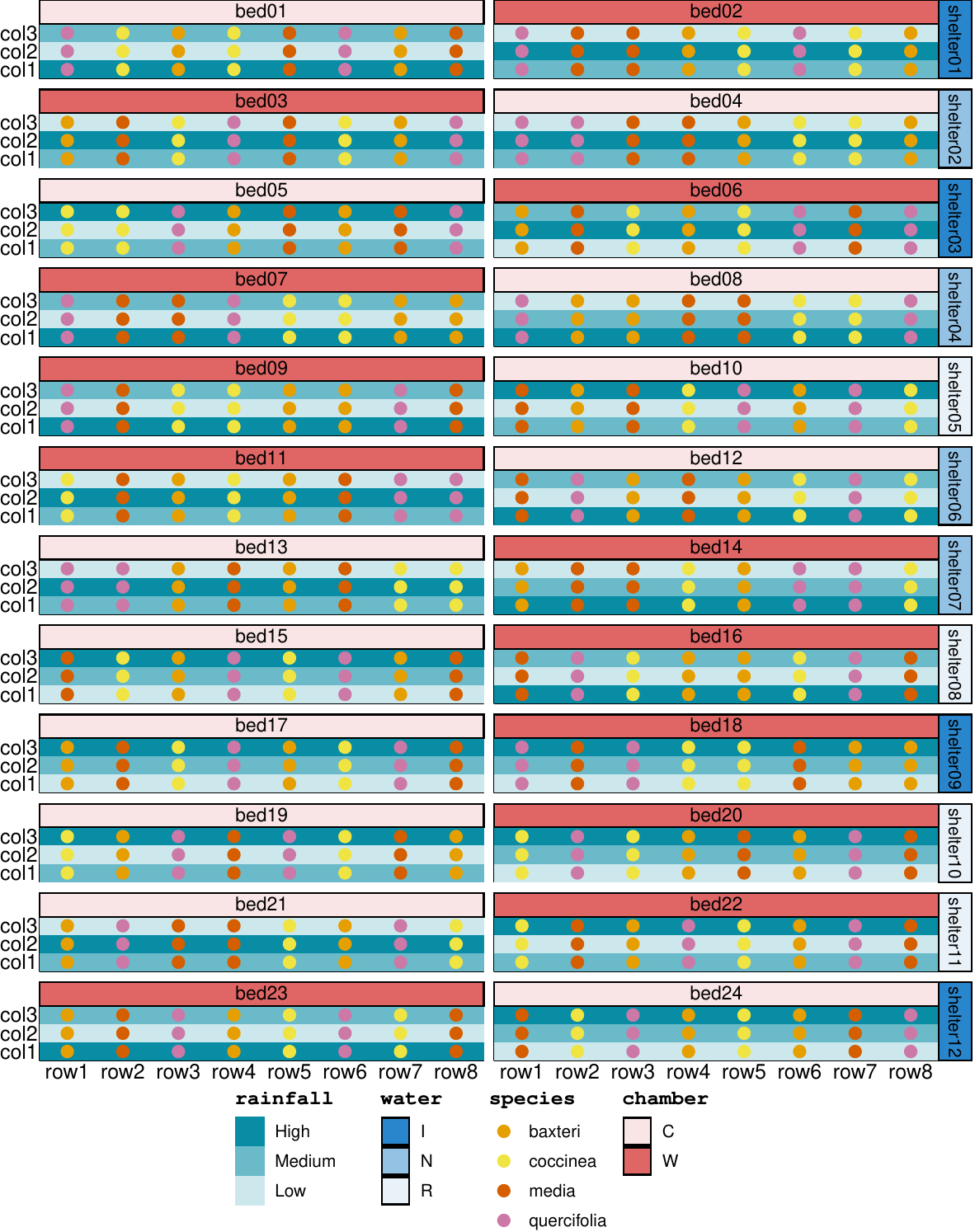}

}

\caption{\label{fig-garden-design1}The generated experimental design in
\texttt{garden\ design1}.}

\end{figure}

Based on the description, I inferred the following mapping between the
treatment factors to unit factors. The default treatment assignment uses
the nesting structures to define the constraint and allocates using the
``random'' order. A random generation of this design is shown in
Figure~\ref{fig-garden-design1}.

\begin{Shaded}
\begin{Highlighting}[]
\NormalTok{garden\_design1 }\OtherTok{\textless{}{-}}\NormalTok{ (garden\_units1 }\SpecialCharTok{+}\NormalTok{ garden\_trts1) }\SpecialCharTok{\%\textgreater{}\%} 
  \FunctionTok{allot\_table}\NormalTok{(water }\SpecialCharTok{\textasciitilde{}}\NormalTok{ shelter,}
\NormalTok{              chamber }\SpecialCharTok{\textasciitilde{}}\NormalTok{ bed,}
\NormalTok{              species }\SpecialCharTok{\textasciitilde{}}\NormalTok{ row,}
\NormalTok{              rainfall }\SpecialCharTok{\textasciitilde{}}\NormalTok{ col,}
              \AttributeTok{label\_nested =} \FunctionTok{c}\NormalTok{(row, col), }
              \AttributeTok{seed =} \DecValTok{2023}\NormalTok{)}
\end{Highlighting}
\end{Shaded}

The design above does not necessarily ensure that ``the same 12
populations always occurred together in a set of four rows''. This split
by four rows sounded like the bed was blocked into two parts across the
rows. Below, we create a block factor with two levels in each bed, then
the block is assigned to rows in a systematic manner.

\begin{Shaded}
\begin{Highlighting}[]
\NormalTok{garden\_units2 }\OtherTok{\textless{}{-}}\NormalTok{ garden\_units1 }\SpecialCharTok{\%\textgreater{}\%} 
  \FunctionTok{set\_units}\NormalTok{(}\AttributeTok{block =} \FunctionTok{nested\_in}\NormalTok{(bed, }\DecValTok{2}\NormalTok{)) }\SpecialCharTok{\%\textgreater{}\%} 
  \FunctionTok{allot\_units}\NormalTok{(block }\SpecialCharTok{\textasciitilde{}}\NormalTok{ row) }\SpecialCharTok{\%\textgreater{}\%} 
  \FunctionTok{assign\_units}\NormalTok{(}\AttributeTok{order =} \StringTok{"systematic{-}slowest"}\NormalTok{)}
\end{Highlighting}
\end{Shaded}

Table A1 in the supplementary material in Cochrane et al. (2015)
provides a list of the population IDs. Upon inspection of these IDs, it
became clear that they were encoded by species, rainfall cline, and
replicate number (e.g., ``Bm H1'', ``Bm H2'', and ``Bm L1'' are
population of species \emph{Banksia media} from site with mean annual
precipitation of 574, 557, and 301, respectively). There were two
replicates for each species and the rainfall cline.

\begin{Shaded}
\begin{Highlighting}[]
\NormalTok{garden\_trts2 }\OtherTok{\textless{}{-}}\NormalTok{ garden\_trts1 }\SpecialCharTok{\%\textgreater{}\%} 
  \FunctionTok{set\_trts}\NormalTok{(}\AttributeTok{rep =} \DecValTok{1}\SpecialCharTok{:}\DecValTok{2}\NormalTok{)}
\end{Highlighting}
\end{Shaded}

\begin{figure}

\centering{

\includegraphics{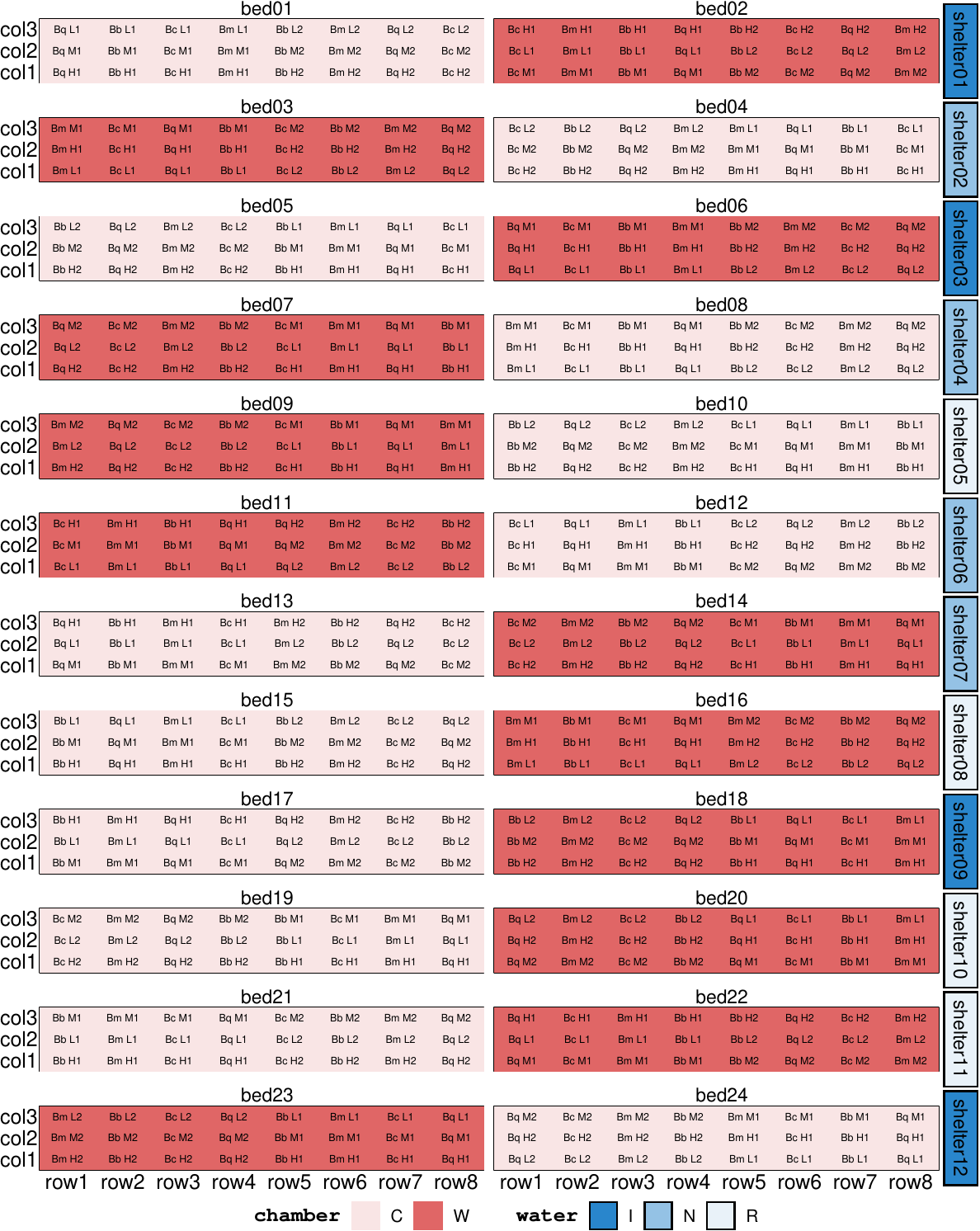}

}

\caption{\label{fig-garden-design2}The generated experimental design in
\texttt{garden\ design2}. The text label in the tile correpond to the
population ID in the supplementary material in Cochrane et al. (2015).}

\end{figure}

Although not specified in the description, the replicate number was
likely allotted to the blocks. The population is then determined from
the species, rainfall, and replication number assigned. This assignment
ensures that ``the same 12 populations always occurred together in a set
of four rows''. The resulting design, shown in
Figure~\ref{fig-garden-design2}, now aligns with all of the text
description.

\begin{Shaded}
\begin{Highlighting}[]
\NormalTok{garden\_design2 }\OtherTok{\textless{}{-}}\NormalTok{ (garden\_units2 }\SpecialCharTok{+}\NormalTok{ garden\_trts2) }\SpecialCharTok{\%\textgreater{}\%}
  \FunctionTok{allot\_table}\NormalTok{(water }\SpecialCharTok{\textasciitilde{}}\NormalTok{ shelter,}
\NormalTok{              chamber }\SpecialCharTok{\textasciitilde{}}\NormalTok{ bed,}
\NormalTok{              species }\SpecialCharTok{\textasciitilde{}}\NormalTok{ row,}
\NormalTok{              rainfall }\SpecialCharTok{\textasciitilde{}}\NormalTok{ col,}
\NormalTok{              rep }\SpecialCharTok{\textasciitilde{}}\NormalTok{ block, }\CommentTok{\# The missing allotment in garden\_design1}
              \AttributeTok{label\_nested =} \FunctionTok{c}\NormalTok{(col, row),}
              \AttributeTok{seed =} \DecValTok{2023}\NormalTok{) }\SpecialCharTok{\%\textgreater{}\%} 
\NormalTok{  dplyr}\SpecialCharTok{::}\FunctionTok{mutate}\NormalTok{(}\AttributeTok{population =} \FunctionTok{paste0}\NormalTok{(}\StringTok{"B"}\NormalTok{, }\FunctionTok{substr}\NormalTok{(species, }\DecValTok{1}\NormalTok{, }\DecValTok{1}\NormalTok{), }\StringTok{" "}\NormalTok{, }
                                    \FunctionTok{substr}\NormalTok{(rainfall, }\DecValTok{1}\NormalTok{, }\DecValTok{1}\NormalTok{), rep))}
\end{Highlighting}
\end{Shaded}

\subsection{Evaluation of Japanese compositions}\label{sec-composition}

E. R. Williams et al. (2021) describe a study on the evaluation of
Japanese language compositions written by Japanese beginners by
assessors of different backgrounds. The background of this experiment is
given in Imaki (2014), with a reduced summary provided below.

\begin{tcolorbox}[enhanced jigsaw, breakable, opacitybacktitle=0.6, leftrule=.75mm, colback=white, opacityback=0, colbacktitle=quarto-callout-note-color!10!white, title=\textcolor{quarto-callout-note-color}{\faInfo}\hspace{0.5em}{Example 3: Evaluation of Japanese composition}, bottomrule=.15mm, arc=.35mm, coltitle=black, bottomtitle=1mm, toprule=.15mm, toptitle=1mm, left=2mm, titlerule=0mm, colframe=quarto-callout-note-color-frame, rightrule=.15mm]

Ten compositions (C1--C10) were conditionally drawn from the Taiyaku
Database of the National Institute of the Japanese Language on a topic
about ``special event in my country''. The compositions were selected
from writers who were Japanese beginners (studied Japanese for less than
300 hours), with a final selection consisting of three compositions each
from Singapore, Brazil, Finland, and a single composition from Cambodia.
The selected compositions have an average of 600 characters.

The study investigated the ratings from people of three different
backgrounds: native Japanese-language teachers (NT), non-native Japanese
language teachers (NNT), and native Japanese who are not language
teachers (NG). While it was intended to involve 10 raters from each
category, availability resulted in 10, 8, and 11 raters, respectively.

Each rater was asked to evaluate the 10 compositions in four aspects
(accuracy, structure and form, context, and richness) plus an overall
score. Each rater was asked to rate the composition in a particular
order (but different order across raters). The raters were given the
option of taking breaks during the assessment process.

Possible carry-over effect was accounted for by using a set of three
different \(10 \times 10\) Williams Latin squares design (E. Williams
1949) where the columns represent the raters and the rows represents the
order of composition presented. As there were 29 raters, one column was
not used.

\end{tcolorbox}

From the description of the experiment, I inferred the following
structure.

\begin{Shaded}
\begin{Highlighting}[]
\NormalTok{composition\_str }\OtherTok{\textless{}{-}} \FunctionTok{design}\NormalTok{(}\StringTok{"Japanese composition"}\NormalTok{) }\SpecialCharTok{\%\textgreater{}\%} 
  \FunctionTok{set\_units}\NormalTok{(}\AttributeTok{background =} \FunctionTok{c}\NormalTok{(}\StringTok{"NT"}\NormalTok{, }\StringTok{"NNT"}\NormalTok{, }\StringTok{"NG"}\NormalTok{),}
            \AttributeTok{rater =} \FunctionTok{nested\_in}\NormalTok{(background, }
                              \StringTok{"NT"} \SpecialCharTok{\textasciitilde{}} \DecValTok{10}\NormalTok{,}
                              \StringTok{"NNT"} \SpecialCharTok{\textasciitilde{}} \DecValTok{8}\NormalTok{,}
                              \StringTok{"NG"} \SpecialCharTok{\textasciitilde{}} \DecValTok{11}\NormalTok{),}
            \AttributeTok{order =} \DecValTok{1}\SpecialCharTok{:}\DecValTok{10}\NormalTok{,}
            \AttributeTok{assess =} \FunctionTok{crossed\_by}\NormalTok{(rater, order)) }\SpecialCharTok{\%\textgreater{}\%} 
  \FunctionTok{set\_trts}\NormalTok{(}\AttributeTok{composition =} \FunctionTok{paste0}\NormalTok{(}\StringTok{"C"}\NormalTok{, }\DecValTok{1}\SpecialCharTok{:}\DecValTok{10}\NormalTok{)) }\SpecialCharTok{\%\textgreater{}\%} 
  \FunctionTok{set\_rcrds\_of}\NormalTok{(}\AttributeTok{assess =} \FunctionTok{c}\NormalTok{(}\StringTok{"accuracy"}\NormalTok{, }\StringTok{"structure"}\NormalTok{, }\StringTok{"context"}\NormalTok{, }
                          \StringTok{"richness"}\NormalTok{, }\StringTok{"overall"}\NormalTok{)) }\SpecialCharTok{\%\textgreater{}\%} 
  \FunctionTok{allot\_trts}\NormalTok{(composition }\SpecialCharTok{\textasciitilde{}}\NormalTok{ assess)}
\end{Highlighting}
\end{Shaded}

The algorithms in {edibble} does not contain an algorithm to generate a
Williams Latin square design (E. Williams 1949), thus we require a
custom treatment assignment order algorithm. The {crossdes} package
(Sailer 2022) has the function {williams} to generate the Williams Latin
squares design; we leverage this function in the algorithm.

In {edibble}, a custom treatment assignment order algorithm is
implemented as an S3 generic function, {order\_trts}, within
{assign\_trts}. A new algorithm is created by writing an S3 method as a
function with an argument that takes in the name of the algorithm,
treatments table, units table, any constraint on the unit that the
treatment is applied, and other objects (such as the provenance object
that contains the entire graph structure). Suppose we name our new
algorithm as ``williams'' then we can write the S3 method
{order\_trts.williams} as follows.

\begin{Shaded}
\begin{Highlighting}[]
\NormalTok{order\_trts.williams }\OtherTok{\textless{}{-}} \ControlFlowTok{function}\NormalTok{(name, trts\_table, units\_table, constrain, ...) \{}
  \CommentTok{\# the experimental unit must be a unit crossed by two unit factors}
  \FunctionTok{stopifnot}\NormalTok{(}\FunctionTok{length}\NormalTok{(constrain)}\SpecialCharTok{==}\DecValTok{2}\NormalTok{)}
\NormalTok{  ntrts }\OtherTok{\textless{}{-}} \FunctionTok{nrow}\NormalTok{(trts\_table)}
  
  \CommentTok{\# assumes unit is ordered correctly}
\NormalTok{  units }\OtherTok{\textless{}{-}} \FunctionTok{lapply}\NormalTok{(units\_table[}\FunctionTok{as.character}\NormalTok{(constrain)], unique) }
\NormalTok{  nunits }\OtherTok{\textless{}{-}} \FunctionTok{sapply}\NormalTok{(units, length)}
  
  \CommentTok{\# smaller unit is assumed to be row, larger unit is the column}
\NormalTok{  unit\_row }\OtherTok{\textless{}{-}} \FunctionTok{as.character}\NormalTok{(constrain[}\FunctionTok{order}\NormalTok{(nunits)}\SpecialCharTok{==}\DecValTok{1}\NormalTok{])}
\NormalTok{  unit\_col }\OtherTok{\textless{}{-}} \FunctionTok{as.character}\NormalTok{(constrain[}\FunctionTok{order}\NormalTok{(nunits)}\SpecialCharTok{==}\DecValTok{2}\NormalTok{])}
  
  \CommentTok{\# the row unit must be exactly equal to the number of treatments}
  \FunctionTok{stopifnot}\NormalTok{(}\FunctionTok{sort}\NormalTok{(nunits)[}\DecValTok{1}\NormalTok{] }\SpecialCharTok{==}\NormalTok{ ntrts)}
  
  \CommentTok{\# generate the William\textquotesingle{}s Latin square design}
\NormalTok{  williams\_design }\OtherTok{\textless{}{-}} \FunctionTok{t}\NormalTok{(crossdes}\SpecialCharTok{::}\FunctionTok{williams}\NormalTok{(ntrts))}
  
  \CommentTok{\# number of times to replicate William\textquotesingle{}s design}
\NormalTok{  nwilliams }\OtherTok{\textless{}{-}} \FunctionTok{ceiling}\NormalTok{(}\FunctionTok{sort}\NormalTok{(nunits)[}\DecValTok{2}\NormalTok{]}\SpecialCharTok{/}\NormalTok{ntrts)}
  
  \CommentTok{\# randomise treatment for each replicate}
\NormalTok{  out }\OtherTok{\textless{}{-}} \FunctionTok{do.call}\NormalTok{(rbind, }\FunctionTok{lapply}\NormalTok{(}\DecValTok{1}\SpecialCharTok{:}\NormalTok{nwilliams, }\ControlFlowTok{function}\NormalTok{(i) \{}
    \FunctionTok{data.frame}\NormalTok{(}\AttributeTok{..trt.. =} \FunctionTok{sample}\NormalTok{(}\DecValTok{1}\SpecialCharTok{:}\NormalTok{ntrts)[}\FunctionTok{as.vector}\NormalTok{(williams\_design)],}
               \AttributeTok{..row.. =} \FunctionTok{as.vector}\NormalTok{(}\FunctionTok{row}\NormalTok{(williams\_design)),}
               \AttributeTok{..col.. =} \FunctionTok{as.vector}\NormalTok{(}\FunctionTok{col}\NormalTok{(williams\_design)) }\SpecialCharTok{+}\NormalTok{ (i }\SpecialCharTok{{-}} \DecValTok{1}\NormalTok{) }\SpecialCharTok{*}\NormalTok{ ntrts)}
\NormalTok{  \})) }\SpecialCharTok{\%\textgreater{}\%} 
    \FunctionTok{subset}\NormalTok{(..col.. }\SpecialCharTok{\textless{}=} \FunctionTok{sort}\NormalTok{(nunits)[}\DecValTok{2}\NormalTok{])}
  
  \CommentTok{\# translate row and column to their unit ids}
\NormalTok{  out[[unit\_row]] }\OtherTok{\textless{}{-}}\NormalTok{ units[[unit\_row]][out}\SpecialCharTok{$}\NormalTok{..row..]}
\NormalTok{  out[[unit\_col]] }\OtherTok{\textless{}{-}}\NormalTok{ units[[unit\_col]][out}\SpecialCharTok{$}\NormalTok{..col..]}
  
  \CommentTok{\# combine and return the allocated treatment id vector }
  \CommentTok{\# ensuring the treatment order matches the units\_table}
\NormalTok{  units\_table}\SpecialCharTok{$}\NormalTok{..id.. }\OtherTok{\textless{}{-}} \DecValTok{1}\SpecialCharTok{:}\FunctionTok{nrow}\NormalTok{(units\_table)}
\NormalTok{  out }\OtherTok{\textless{}{-}} \FunctionTok{merge}\NormalTok{(units\_table, out)}
\NormalTok{  out[}\FunctionTok{order}\NormalTok{(out}\SpecialCharTok{$}\NormalTok{..id..), }\StringTok{"..trt.."}\NormalTok{, drop }\OtherTok{=} \ConstantTok{TRUE}\NormalTok{]}
\NormalTok{\}}
\end{Highlighting}
\end{Shaded}

\begin{figure}

\centering{

\includegraphics{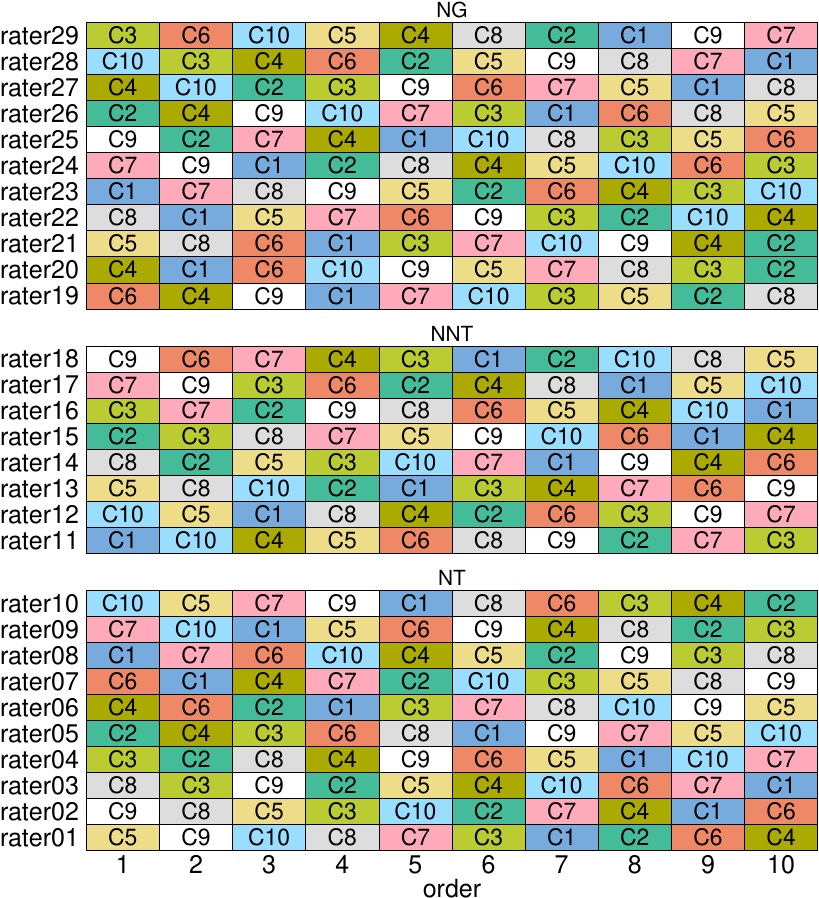}

}

\caption{\label{fig-composition}An experimental design for the
evaluation of Japanese compositions. The plot shows the order in which
the ten composition was evaluated by each rater with the rater grouped
by their background.}

\end{figure}

The above algorithm is used by setting the argument
\texttt{order\ =\ "williams"}. The above function contains checks to
ensure the experimental structure is as expected before proceeding to
the allocation. This check makes it clear what the expected experimental
structure is for the selected ordering algorithm. The resulting design
is illustrated in Figure~\ref{fig-composition}.

\begin{Shaded}
\begin{Highlighting}[]
\NormalTok{composition\_design }\OtherTok{\textless{}{-}}\NormalTok{ composition\_str }\SpecialCharTok{\%\textgreater{}\%} 
  \FunctionTok{assign\_trts}\NormalTok{(}\AttributeTok{order =} \StringTok{"williams"}\NormalTok{, }\AttributeTok{seed =} \DecValTok{2023}\NormalTok{) }\SpecialCharTok{\%\textgreater{}\%} 
  \FunctionTok{serve\_table}\NormalTok{()}
\end{Highlighting}
\end{Shaded}

\subsection{Menu Designs}\label{menu-designs}

While selecting a design from a set of ``named'' or ``recipe'' designs
are not encouraged, there are cases when it is convenient to refer to
these. In {edibble}, functions that prefix with \texttt{menu\_} generate
the full recipe code in their basic terms. For example, a completely
randomised design (CRD) can be generated as:

\begin{Shaded}
\begin{Highlighting}[]
\FunctionTok{menu\_crd}\NormalTok{(}\AttributeTok{n =} \DecValTok{10}\NormalTok{, }\AttributeTok{t =} \DecValTok{2}\NormalTok{, }\AttributeTok{seed =} \DecValTok{1}\NormalTok{)}
\end{Highlighting}
\end{Shaded}

\begin{verbatim}
design("Completely Randomised Design") %>%
  set_units(unit = 10) %>%
  set_trts(trt = 2) %>%
  allot_trts(trt ~ unit) %>%
  assign_trts("random", seed = 1) %>%
  serve_table()
\end{verbatim}

If no argument is supplied, then the code automatically generates a
random number:

\begin{Shaded}
\begin{Highlighting}[]
\FunctionTok{menu\_crd}\NormalTok{()}
\end{Highlighting}
\end{Shaded}

\begin{verbatim}
design("Completely Randomised Design") %>%
  set_units(unit = 21) %>%
  set_trts(trt = 10) %>%
  allot_trts(trt ~ unit) %>%
  assign_trts("random", seed = 871) %>%
  serve_table()
\end{verbatim}

The above output is a code print out designed so that users can copy and
paste the output and easily modify the numbers or factor names to their
context. If you would like to actually run the code from the menu
output, then the {takeout} function allows you to do this easily:

\begin{Shaded}
\begin{Highlighting}[]
\FunctionTok{takeout}\NormalTok{(}\FunctionTok{menu\_crd}\NormalTok{(}\AttributeTok{n =} \DecValTok{4}\NormalTok{, }\AttributeTok{t =} \DecValTok{2}\NormalTok{))}
\end{Highlighting}
\end{Shaded}

\begin{verbatim}
design("Completely Randomised Design") %>%
  set_units(unit = 4) %>%
  set_trts(trt = 2) %>%
  allot_trts(trt ~ unit) %>%
  assign_trts("random", seed = 798) %>%
  serve_table() 

# Completely Randomised Design 
# An edibble: 4 x 2
    unit    trt
* <U(4)> <T(2)>
   <chr>  <chr>
1  unit1   trt1
2  unit2   trt1
3  unit3   trt2
4  unit4   trt2
\end{verbatim}

If no argument is supplied for {takeout}, a random recipe design is
selected from the menu. The full menu can be seen from {scan\_menu}.

\begin{Shaded}
\begin{Highlighting}[]
\FunctionTok{scan\_menu}\NormalTok{()}
\end{Highlighting}
\end{Shaded}

\begin{verbatim}
# A tibble: 10 x 4
   package name         args                 name_full                          
   <chr>   <chr>        <chr>                <chr>                              
 1 edibble bibd         t, k, r, seed        Balanced Incomplete Block Design   
 2 edibble crd          t, n, r, seed        Completely Randomised Design       
 3 edibble factorial    trt, r, design, seed Factorial Design                   
 4 edibble graeco       t, seed              Graeco-Latin Square Design         
 5 edibble hyper_graeco t, seed              Hyper-Graeco-Latin Square Design   
 6 edibble lsd          t, seed              Latin Square Design                
 7 edibble rcbd         t, r, seed           Randomised Complete Block Design   
 8 edibble split        t1, t2, r, seed      Split-Plot Design, Split-Unit Desi~
 9 edibble strip        t1, t2, r, seed      Strip-Plot Design, Strip-Unit Desi~
10 edibble youden       nc, t, seed          Youden Square Design               
\end{verbatim}

\section{Comparisons with other packages}\label{sec-contrast}

There are currently 103 R-packages in the CRAN Task View of Design of
Experiments and Analysis of Experimental Data (as of 2023-11-16). The
list is a comprehensive, although not exhaustive, list of R-packages
that can construct experimental designs, and some included in the list
have an analytical focus or serve as a repository of experimental data
(Tanaka and Amaliah 2022).

The packages used to construct experimental designs can be broadly
categorised into three approaches: specialist, recipe or general
purpose. The first type is exclusive to a domain or has a limited
experimental structure. The second often makes use of named experimental
designs, e.g., the {agricolae} package (de Mendiburu 2023) specifies a
Latin square design and a split-plot design by the function {design.lsd}
and {design.split}, respectively. Alternatively, a package can contain a
catalogue of pre-optimised designs, e.g., {FrF2} (Gromping 2014). The
general purpose approach can construct a more flexible range of
experimental designs using either a randomisation or optimisation
methods. E.g. {AlgDesign} package (Wheeler 2022) includes a {optFederov}
function finds an optimal design using Federov's exchange algorithm
under certain optimal criterion.

A specialist or recipe approach clearly does not easily allow users to
deviate away from a named experimental design. A general purpose
approach seems to offer the most flexibility but it often assumes a
\emph{mise en place}; in other words, the ingredients are prepared
before ``cooking''. A prime example of this is optimal or model-based
design, such as in {AlgDesign}, which requires the initial data
structure and specification of the algorithm type and/or model. There
are often helper functions to obtain the required initial data structure
(such as {gen.factorial} in {AlgDesign}).

All of the above mentioned packages have an emphasis on the algorithm to
generate the experimental design. On the other hand, {edibble} package
serves as a holistic approach in which segments of the process are
modularised. The algorithms, catalogues, or recipes can be integrated
into the system as shown in Section~\ref{sec-composition}; in this
sense, the system is largely complementary to existing systems and
offers fundamentally different functionalities. A prime example is the
downstream usage of record factors to enter data validation rules (see
Section~\ref{sec-rcrds}), which are not offered in the other
experimental design packages.

\section{Discussion}\label{sec-conclude}

This paper showcases the capabilities of the {edibble} package, an
implementation of the grammar of experimental designs system by Tanaka
(2023) in the {R} language. While it is conceivable to implement the
system in other programming languages, the concrete implementation as an
{R} package should aid in a faster adoption by the scientific community
to make their experiments transparent.

The demonstrated downstream benefits of {edibble} include the ability to
specify the record (including responses) of unit factors ({set\_rcrds}),
and the expected values these records ({expect\_rcrds}) can take, which
in turn can be encoded as data validation rules in the exported data
({export\_design}). These functionalities are also leveraged in the
simulation of record factors ({simulate\_rcrds}), including in a lazy,
automated fashion ({autofill\_rcrds}). These functionalities aid in
better planning and management, and consequently, in the workflow for
constructing the experimental design.

Software packages for experimental designs often focus on the
algorithmic components of the design. In contrast, {edibble} package
reframes the specification of the experimental design by conceiving
experimental designs as a mutable object that can be progressively
built. This reframing allows separation in the process of specifying the
experimental structure and the assignment algorithm. The benefit of this
approach is that users can easily consider various assignment algorithms
without respecifying their structure. This approach also aligns with the
common mantra that a good experimental design considers the structure
first rather than just selecting a recipe design from the menu (Bailey
2008).

Mead, Gilmour, and Mead (2012) attest Procrustean designs that
compromise the experimental objective are still common. The {edibble}
package takes a holistic approach that aims to encourage users to think
of the experimental structure. By engaging user cognition, this can
potentially translate into a better practice of experimental design.

The {edibble} package can also aid in the teaching of experimental
design. Smucker et al. (2023) found from their survey of experimental
design courses that nearly all cover randomisation, replication,
blocking design principles, and multi-factor experiments, and the use of
software for both instruction and assessment is ubiquitous. By adhering
closely to the principles of {tidyverse} (Wickham et al. 2019), which
has widespread adoption in modern day practice and teaching
(Çetinkaya-Rundel et al. 2022), {edibble} can integrate more easily in
the workflow of everyday users or learners by leveraging their knowledge
in a familiar system.

The {edibble} package is fairly flexible in constructing various fixed
experimental structures, and it is fairly easy to wrap existing
experimental design algorithms into the system, as shown in
Section~\ref{sec-composition}. The ability to incorporate other
algorithms into the system provides an opportunity for users to select
the most appropriate algorithm for their structure. However, there is no
guarantee that the generated experimental design will be appropriate or
optimal. Afterall, a tool is only as effective as the skillfulness of
the hands that wield it. The {edibble} package aims to cognitively
engage the user with the hope that possible broader concerns in the
experimental design are detected prior to the execution of the
experiment, but ultimately, the user should ensure to perform their own
diagnostic checks that the experimental design output is appropriate for
their own objectives.

\section*{Computational details}\label{computational-details}
\addcontentsline{toc}{section}{Computational details}

The paper was written using {Quarto} version 1.4.502 with {knitr}
version 1.45, Pandoc version 3.1.1 and {R} version 4.3.1 (2023-06-16).
Some of the figures were drawn using {ggplot2} version 3.4.4. The code
uses {edibble} version 1.1.0.

\section*{Acknowledgements}\label{acknowledgements}
\addcontentsline{toc}{section}{Acknowledgements}

I thank Francis Hui and many others that supported the idea and provided
feedback on the system.

\section*{References}\label{references}
\addcontentsline{toc}{section}{References}

\phantomsection\label{refs}

\end{document}